\documentclass[12pt,preprint]{aastex}
\usepackage{graphicx}

\newcommand{\mn}{\ion{Mn}{2}}
\newcommand{\ca}{\ion{Ca}{2}}
\newcommand{\mg}{\ion{Mg}{2}}

\newcommand{\kms}{km s$^{-1}$}
\newcommand{\mum}{$\mu$m}

\received{2002 September 12}
\begin{document} 

\title{Near Infrared Spectra of Type Ia Supernovae}

\shorttitle{Near Infrared Spectra of Type Ia Supernovae}
\shortauthors{Marion et al.}

\author{G. H. Marion\altaffilmark{1,2} \& P. H\"oflich}
\affil{Astronomy Department, University of Texas at Austin, 
       Austin, TX 78712, USA}

\author{W. D. Vacca\altaffilmark{2,3}}
\affil{Max-Planck-Institut fuer extraterrestrische Physik\\
       Postfach 1312, D-85741 Garching, Germany}
\and

\author{J.C. Wheeler} \affil{The University of Texas at Austin,
Department of Astronomy,  1 University Station C1400,  Austin, TX
78712-0259, USA}

\altaffiltext{1}{email: {\tt hman@astro.as.utexas.edu}}

\altaffiltext{2}{Visiting Astronomer at the Infrared Telescope
Facility, which is operated by the University of Hawaii under contract
from the National Aeronautics and Space Administration.}

\altaffiltext{3}{current address: The Dept. of Astronomy, 601
Campbell Hall, Univ of California at Berkeley, Berkeley, CA 94720}

\begin{abstract}

We report near infrared (NIR) spectroscopic observations of twelve
``Branch-normal'' Type Ia supernovae (SNe Ia) which cover  the
wavelength region from 0.8-2.5 \mum.  Our sample more than doubles the
number of SNe Ia with published NIR spectra within three weeks of
maximum light.  The epochs of observation range from thirteen days
before maximum light to eighteen days after maximum light.  A detailed
model for a Type Ia supernovae is used to identify spectral features.
The Doppler shifts of lines are measured to obtain the velocity and,
thus, the radial distribution of elements.

The NIR is an extremely useful tool to probe the chemical structure in
the layers of SNe Ia ejecta.  This wavelength region is optimal for
examining certain products of the SNe Ia explosion that may be blended
or obscured in other spectral regions.  We identify spectral features
from \mg, \ca, \ion{Si}{2}, \ion{Fe}{2}, \ion{Co}{2},  \ion{Ni}{2},
and possibly \mn.   We find no indications for hydrogen, helium or
carbon in the spectra.
The spectral features reveal important clues about the physical
characteristics of SNe Ia.  We use the features to derive upper limits
for the amount of unburned matter, to identify the transition regions
from explosive carbon to oxygen burning and from partial to complete
silicon burning, and to estimate the level of mixing during and after
the explosion.

Elements synthesized in the outer layers during the explosion appear
to remain in distinct layers.  That provides strong evidence for the
presence of a detonation phase during the explosion as it occurs in
delayed detonation or merger models.  \mg\ velocities are found to
exceed 11,000 to 15,000 \kms\ depending on the individual SNe Ia. That
result suggests that burning during the explosion reaches the
outermost layers of the progenitor and limits the amount of unburned
material to less than 10\% of the mass of the progenitor.  Small
residuals of unburned material are predicted by delayed detonation
models but are inconsistent with pure deflagration or merger models.
Differences in the spectra of the individual SNe Ia demonstrate the
variety of these events.

\end{abstract}


\section {Introduction}

Type Ia supernovae (SNe Ia) are excellent distance indicators due to
their brightness and apparent homogeneity.   The high quality of
SNe~Ia observations has allowed good estimates of the Hubble constant
\cite{hamuy96, hk96, Muller94, nugent97, Riess95, phillips99}, and
provided strong evidence for the existence of ``dark energy''
\cite{Riess98a, Perl_99}.  The quest for the nature of the dark energy
requires even higher accuracy \cite{WA2001}.  Accurate determinations
of cosmological parameters will require a more thorough understanding
of the properties (e.g., composition and metallicity) of the
progenitors of SNe Ia, as well as of the nature and time-evolution of
the explosion mechanism.

When we observe a supernova, we are collecting light that has been
emitted from a rapidly expanding envelope.  As expansion of the
explosion products reduces the density, the photosphere recedes toward
the center in mass space.  Deeper layers of the supernova become
visible.  Because SNe Ia ejecta expand homologously soon after the
explosion, we can measure the chemical composition of the layers of
explosion products by observing the Doppler velocities of elements
within the layers.

The observed characteristics of SNe Ia strongly suggest that these
events are the thermonuclear explosions of carbon/oxygen (C/O) White
Dwarf (WD) stars \cite{hoyle60}.  Within this general picture, two
classes of models are most likely: 1) The explosion of a CO-WD, with
mass close to the Chandrasekhar limiting mass ($M_{Ch}\approx
1.4M_{\odot}$).  The progenitor accretes mass through  Roche-lobe
overflow from an evolving companion star \cite{whelan73} and the
explosion is triggered by compressional heating near the WD center.
2) The explosion of a rotating configuration formed from the merger
of two low-mass  WDs, caused by the loss of angular momentum through
gravitational radiation \cite{webbink84, iben84, paczynski85}.
Observations of optical light curves and spectra favor the first model
for the majority of SNe Ia.

For $M_{Ch}$ models, it is believed that the burning front begins as a
subsonic deflagration, but the time evolution of the burning front
remains an open question. It is not known whether the deflagration
front burns through the entire WD \cite{nomoto84}, or alternatively,
makes a transition into a supersonic detonation mode as suggested in
the delayed detonation (DD) model \cite{khokhlov91, woosley94,
yamaoka92}.  Although  DD models have been found to reproduce the
optical and infrared light curves and spectra of ``typical'' SNe Ia
reasonably well \cite{h95, hk96, fisher98, nugent97, lentz01}, 
the calculations assume spherical geometry, and the propagation of the
deflagration burning front is parameterized.


While it is preferable to model SNe Ia using full 3-D calculations, at
the present time there are limitations to that approach.  Although the
propagation of a detonation front is well understood, the description
of the deflagration front and the deflagration to detonation
transition (DDT) pose problems for the models.  The current state of
the art in 3-D modeling of the deflagration burning front is limited
to a description of the large scale instabilities that dominate only
during the early phase of the explosion \cite{livne93, khokhlov95,
reinecke99, lisewski00, khokhlov01}.  These models do not resolve
small scale instabilities, do not use realistic WD structures and do
not include instabilities due to differential rotation or pre-existing
velocity fields all of which are thought to be important for the late
stages of  deflagration and the transition from a deflagration to a
detonation  front.  For a discussion, see (H\"oflich et al. 2003; and
references therein).
The current 3D calculations do  not show a DDT but remain in the
deflagration burning phase.   These models produce structures with a
significant fraction of the WD ($\approx 0.4 M_\odot$) remaining
unburned in the outer layers.  Despite some uncertainties, a common
feature of pure deflagration models is the complicated morphology of
the burning front and the burning products.  Plumes of burned material
fill a significant fraction of the WD, and unburned or
partially-burned material can be seen near the center.  Thus iron-rich
elements are not confined to the central region as in 1-D models.
While the expansion of the envelope becomes almost spherical, the
inhomogeneous chemical structure will fill about 50 to 70\% of the
volume of the expanding envelope.  This prediction that both burned
and unburned material will be present throughout the envelope is not
what is observed.  Our data suggest that there is a distinct layering
of the chemical products.  In pure deflagration models, a homogeneous,
unburned layer of $\approx$ 0.2 to 0.3 $M_\odot$ remains on top
because this region quickly attains velocities in excess of the sound
speed and, consequently, cannot be reached by the rising plumes.  If a
DDT occurs at densities required to reproduce normal-bright SNe Ia,
most of the unburned fuel in these outer regions will be burned to
intermediate mass or iron-group elements during the detonation
phase. That would eliminate the chemical inhomogeneities
but is inconsistent with observations of intermediate mass elements in
the spectra of SNe Ia.  The study of NIR spectra can help resolve
these contradictions.

The line identifications we present here are the result of detailed
model calculations of the hydrodynamics and nucleosynthesis in SNe Ia
as discussed in Section~\ref{rmod}.  The position of the photosphere
and the line forming region of the supernova is calculated.  Specific
line identifications result from model calculations that predict the
relative line strengths at the epoch of our observations.

Near infrared (NIR) analysis promises to provide significant new
constraints on SNe Ia physics by providing access to information about
progenitor composition and explosion products.  NIR spectra reveal
features from elements such as He, C, O, Mg, and Mn that are
undetectable or obscured by line blending at other wavelengths.  \mg\
(0.922 and 1.0926 \mum) is an indicator of the boundary between
explosive carbon and oxygen burning during the explosion.  \mn\ (0.944
\mum) is a probe of the burning temperature in the region of
incomplete silicon burning.  
The Ca ``infrared triplet'' (0.850, 0.854, 0.866 \mum) is strong
enough to respond to primordial calcium abundances but the \ca\ line
at 1.268 \mum\ will be from freshly synthesized material and indicates
the location of the region with partial Si burning.  \ion{C}{1} (0.93
and 1.13 \mum) is an indicator of the presence of unburned material
from the progenitor WD.  The position and development with time of
blended lines from iron-group elements ($1.6-1.8$ \mum) is an
indicator for the transition from partial to complete silicon burning.
Relative photometric data are available for nine of the SNe in our
sample.  We are able to make comparisons between the decline rate
($\Delta m_{15}$) \cite{phillips99} and our NIR data.  The parameter
$\Delta m_{15}$ is the decline in apparent brightness between the peak
and fifteen days later.  The calibrated brightness at maximum is
unavailable for these data and peak absolute brightness measurements
from other sources contain uncertainties of $\approx \pm0.5$ mag which
makes that parameter unsuitable for comparison.

SNe Ia have been observed extensively at optical and radio wavelengths
but only recently have improvements in detector technology opened the
NIR window to objects as dim as a typical SNe Ia.  Relatively few NIR
spectra have been published: NIR spectroscopic observations obtained
within three weeks of maximum light are available for only  four
normal-bright SNe.  SN 1986G was observed 13 days after maximum light
\cite{Bowers97}.  Six spectra from -8 to +8 days were obtained from SN
1994D \cite{Meikle_96}. Spectra of SN 2000cx were obtained at -8 and
-7 days before maximum by Rudy et al. (2002).  An excellent time
sequence for SN 1999ee from -9 to +42 days was obtained by Hamuy et
al. \cite{hamuy02}.  A similar set of NIR spectra were obtained  by
Gerardy of the subluminous SNe Ia SN 1999by between -4 to +28 days
(H\"oflich et al. 2002).  In addition, NIR spectra obtained more than
25 days after maximum light have been published by Jha (1999), Bowers
(1997), and Hernandez (2000).  Although the data are significant, the
actual number of SNe Ia observed in the NIR remains small. A larger
sample is critical to address the diversity of SNe Ia and, in
particular, the role of pre-conditioning of the WD prior to the
explosive phase.  The explosion depends on the WD initial conditions
which are determined by the progenitor evolution from the main
sequence \cite{umeda00,dominguez00}, during the accretion phase and
during the evolution towards the runaway \cite{nomoto82, garcia95,
hs02}.

We present NIR spectra of twelve SNe Ia which more than doubles the
number of SNe Ia with published NIR observations.  We provide
identification of significant features in the spectra and their
location in the expanding envelope.  We compare the results to a model
of a normal-bright SNe Ia.  Data acquisition and reduction methods are
discussed in \S~\ref{data}.  Techniques used to determine the epoch
and brightness of the supernovae in our sample are described in
\S~\ref{sne}.  Methods of analysis including details of the reference
model for a delayed detonation explosion of a normal-bright Type Ia
supernova are presented in \S~\ref{datanal}.  The spectral features are
identified and discussed in \S~\ref{spectra}.  Implications from the
data for the models are discussed in \S~\ref{models}.  Conclusions are
presented in \S~\ref{conc}.  The observational details of discovery,
identification and our spectroscopic observations for each supernova
are provided in Appendix ~\ref{sn}.

\section {Data Acquisition and Reduction}
\label{data}

Low and medium resolution, NIR spectra from SNe Ia were obtained using
the 3.0 meter telescope at the NASA Infrared Telescope Facility (IRTF)
with the SpeX medium-resolution spectrograph \cite{Rayner98}.  The
SpeX instrument provides single exposure coverage of the wavelength
region from 0.8-2.5 \mum.  Using a grating and prism cross-dispersers,
the spectral resolution is $R=750-2000$ and in single prism mode
$R=150-250$.  SpeX also contains an infrared slit-viewer/guider
covering a 60x60 arcsec field-of-view at 0.12 arcsec/pixel.  
The detectors are a Raytheon 1024x1024 InSb array in the spectrograph
and a Raytheon 512x512 InSb array in the infrared slit-viewer
\cite{Rayner02}.

Thirteen NIR spectra were obtained at the IRTF from twelve SNe Ia at
epochs ranging from thirteen days before maximum light to eighteen
days after maximum light (Fig. \ref{sn13}). Two spectra were obtained
three weeks apart from SN 2001en.  All spectra cover the wavelength
region from 0.8-2.5 \mum.  Six of the spectra in our sample were
obtained using the 0.8 arcsec slit in cross-dispersed (SXD) mode which
gives R=750.   Four spectra were obtained with the 0.5 arcsec slit in
SXD mode at R=1200.  Three of the spectra were obtained using the
single prism, low resolution (LRS) mode with the 0.3 arcsec slit at
R=250.  Table~\ref{obs_table} provides the details for each
observation.

For each set of exposures, care was taken to align the slit to the
parallactic angle.  In a few cases, light contamination due to the
position of the host galaxy made it impossible to achieve the exact
parallactic angle.  However, these observations had relatively low
air-masses (less than 1.5) so the errors due to atmospheric refraction
should not be significant.

The guiding method was determined by target luminosity and weather
conditions.  For SNe brighter than $J\sim15$, the SpeX guider is able
to maintain the centroid of the target in the slit by guiding on the
spill-over flux from the object in the slit.  When the target is not
bright enough to produce sufficient spill-over, we use the IRTF
optical offset guider or the offset method with the SpeX guider on
another object in the field of the SpeX imager, such as the core of
the host galaxy, for the guider reference.

Saturation was not a concern due to the faintness of our
objects, but OH lines are numerous and highly variable in the NIR.  To
avoid an increase in background noise due to poor OH removal when the
pairs were subtracted, we limited the exposure time to 150s per
exposure.   Each set was also  limited to ten exposures so that the
time required to complete each set, including calibration, was within
the time scales of atmospheric variability. The source was nodded along
the slit using an A\_B\_B\_A\_A\_B\_B\_A\_A\_B pattern for a total of
25 minutes integration time.  Calibration images from A0V standard
stars were obtained in a sequence similar to the SN exposures, but
with shorter exposure times.  The standards were selected to be as
near as possible to the time and airmass of the SN.  Each observation
set also included calibration images for flat fielding using an
internal light source, and wavelength calibrations using an Argon lamp
through the slit.

The data were reduced using a package of IDL routines specifically
designed for the reduction of SpeX data (Spextool v. 2.1; Cushing,
Vacca \& Rayner 2002). These routines perform pair subtraction,
flat-fielding, aperture definition, spectral tracing and extraction,
residual sky subtraction, host galaxy subtraction, and wavelength
calibration for data acquired in both the prism mode and the
cross-dispersed mode.  Corrections for telluric absorption were
performed using the extracted spectrum of an A0V star and a specially
designed IDL package developed by Vacca, Cushing, \& Rayner
(Spextool\_Extension; 2002). These routines generate a telluric
correction spectrum by comparing the spectrum of an A0V star, observed
close in time and airmass to the target SN, to a model A0V spectrum
that has been scaled to the observed magnitude, smoothed to the
observed resolution and shifted to the observed radial velocity. The
telluric correction spectrum is then shifted to align the telluric
absorption features seen in the SN spectrum and divided into the
target spectrum.

Spextool removes problems with background irregularities by defining
the aperture width and background level for each subtracted image.
The flux level along the slit includes a positive image of the SN at
the A position and a negative image of the SN at the B position.  The
position, shape, and slope of the background level are due to
sky and instrumental noise, plus contamination from the host galaxy.
The aperture width for each pair is selected to enclose the region of
supernova light within the slit.  
The full width of the supernova in the image is 1.2-1.5'', but optimal
signal to noise ratio in the reduced spectrum was achieved by
narrowing the aperture to 0.7-0.9''.  The narrower aperture eliminates
high noise levels from the wings of the supernova signal.  The
background level in the slit is defined by Spextool using a fit
through four regions, in two pairs, that bracket each
aperture. Typically this is a linear fit, but for noisy spectra a
second or third order fit was more effective.

The Spex Guider was used to obtain J and K band images of the field
surrounding some of our targets in order to estimate levels of
galactic contamination.   We reduced the spectra using background
levels measured photometrically from these images and compared the
results to spectra reduced using Spextool.  We found the energy
distribution for the spectra produced by both methods to be parallel
through the entire wavelength region.   The differences in flux levels
between the spectra produced by the two methods were less than the
noise levels.  We conclude that contamination from the host galaxy
does not influence the slope of the continuum in spectra reduced using
 Spextool.

The spectra are extracted individually for each AB pair of images and
the  
individual multi-order spectra from each set were
combined to produce a single spectrum.  Next, the spectra from
the various sets for each object were combined.  The
orders of the final combined spectra were then merged using a Spextool
routine that allows the orders to be scaled and then averaged in their
mutual overlap regions.

\section {The Supernovae}
\label{sne}

Details of discovery and photometry for each SN in our sample are
compiled in Appendix A.  Observational details may be found in
Tables~\ref{obs_table} and \ref{norm_table}.

\subsection{The Epochs of Observation}
A well-calibrated light curve, sufficiently sampled to clearly define
the maximum brightness will establish the maximum light date to within
a day or two.
W. D. Li and M. Papenkova from the Katzman Automatic Imaging Telescope
\cite{kait} and the University of California at Berkeley have kindly
provided B-band relative light curves for nine SNe covering ten of the
spectra in our sample.  These light curves allow us to define the date
of $B_{max}$ to $\pm 1-2$ days and $\Delta m_{15}$ to $\pm0.2$ mag.
The relative nature of the data do not provide information about peak
brightness.  Epochs used for discussion always refer to the date of
maximum light in V and $V_{max}$ is taken to be $B_{max} + 2$ days.

Light curves for the other SNe in our sample are found in lists of
amateur photometric data, but these data have been collected from many
observers without calibration between the instruments or observers.
Interpretation of such light curves is difficult due to the number of
obvious outliers.  Uncertainties of five to seven days for the date of
maximum light and $\pm 0.5$ mag in observed luminosity are not
uncommon.  For our analysis we use amateur photometric data compiled
by VSnet (The Variable Star Network) and AUDE (Association des
Utilisateurs de D\'{e}tecteurs Electroniques).  A good source for data
from individual SNe is {\sl The Latest Supernova Page}, compiled and
maintained by David Bishop \cite{LSP}.

\subsection {Uniformity of the Sample}

\subsubsection {Absolute Brightness}

All SNe in our sample were identified spectroscopically as Type Ia
using the \ion{Si}{2} feature at $\lambda=6355$ which defines the
class.  The observed maximum luminosities  indicate that all events
are, within uncertainties, normal-bright events.  $M_V$ is computed
for each event in the sample using

\begin{displaymath}
M_V=V_{obs} - 5 \times \log(v_{rec}/H_0) - 25 - A_V
\end{displaymath} 

\noindent{where $H_0=65$ \kms Mpc$^{-1}$ and $A_V=3.1 \times E(B-V)$.}

The results are presented in Table~\ref{norm_table}.  Galactic
foreground reddening data are from Schlegel, Finkbeiner, \& Davis
\cite{Schlegel98} as reported in the NASA/IPAC Extragalactic
Database \cite{NED}.  We do not account for reddening from the host
galaxy.  The NED database also provides the recession velocities of
the host galaxies.  All SNe in our sample may be considered to be in
the Hubble flow since uncertainties in the recession velocity of the
host produce uncertainties in our calculation of $M_V$ that are much
less than
the $\pm 0.5$ mag uncertainties due to the photometric data.
Our calculation of $M_V$ is intended as a demonstration that, within
the uncertainties, our sample contains only normal-bright SNe Ia.  The
results should not be interpreted as a reliable measure of absolute
brightness for these events.

\subsubsection{Empirical Photometric Parameters}

Values for $\Delta m_{15}$ (B-band) have been calculated for ten of
the spectra in our sample.  The values range from $0.8-1.5$ mag and
are listed in Tables~\ref{norm_table} and \ref{v_table}.  $\Delta
m_{15}$ was directly calculated from the photometric data and not by
fitting the data to templates.  The uncertainties of $\approx \pm0.2$
mag provide sufficient resolution to distinguish fast decliners from
slow decliners.  We do not find any correlation between the $\Delta
m_{15}$ parameter and the expansion velocity of explosion products.

\section{Methods for the Data Analysis}
\label{datanal}

Line identifications for the spectral features are based on detailed
calculations of the explosion, the position of the photosphere and the
line forming region of the supernova.  Opacity and optical depth are
calculated as a function of radius (see Figure 8, H\"oflich et
al. 1998).  Together with synthetic spectra, these calculations allow
us to identify features in the observed spectra.  Details of the model
used for this analysis are provided in the following section.

All models that are rich in C and O, such as $M_{Ch}$ or WD mergers,
produce lines with similar excitation values independent of the
details of the particular model.  For example, in the region of
incomplete Si burning, the burning products maintain a similar ratio
of Si to S, Ar, and Ca (see Figure~\ref{logyi}).  Therefore we can
confidently use a reference model for a normal bright Type Ia
supernova to provide line identifications for the spectra presented
here.  Specific line identifications are made by calculating the
relative line strengths at the epoch of our observations.  Many of the
lines we discuss here have been identified in previous work
\cite{Wheeler98, hkt98, PAH02}.  It is possible however, for different
model scenarios (e.g. mergers or sub-Chandrasekhar mass models/Helium
Detonations) to create different layering structures that produce a
similar feature at a given wavelength but due to a different element.
For example, near maximum light, both $M_{Ch}$ and HeDs produce a line
at about 1.05 but which is produced by Mg II and He I, respectively.
Additional line features (or their absence) of the same ion are
critical.  Identifications based on a single line cannot be positively
confirmed.  The \mn\ line at 0.922 \mum\ is an example of an
identification that is consistent with our standard model but is
uncorroborated by a second \mn\ line.  It is a significant advantage
that the large wavelength coverage of our spectra is able to reveal
multiple lines for many elements in the NIR.  The line identifications
for \mg, \ca, \ion{Si}{2}, \ion{Fe}{2}, \ion{Co}{2}, \ion{Ni}{2} are
based on multiple lines in our spectra.

We measure the Doppler shifts of the lines to obtain the expansion
velocities of the elements that produce the features.  Since SNe Ia
begin homologous expansion soon after the explosion, the radial
velocities reveal the the radial distribution of explosion products.

\subsection{The Reference Model for the Line Identification}
\label{rmod}

Calculations of the reference models were performed using our
hydrodynamical radiation transport code HYDRA which includes
hydrodynamics, $\gamma-$ray and low energy photon transport schemes, a
nuclear reaction network and detailed atomic networks to solve the
NLTE rate equations for the atomic level populations (see  H\"oflich
et al. 2002, H\"oflich 2002, and references therein).  The explosion
models, light curves, and synthetic spectra are calculated in a
self-consistent manner. Given the initial structure of the progenitor
and a description of the nuclear burning front, the light curves and
spectra are calculated from the explosion model.

Our reference model is based on a delayed detonation scenario because
of the success of such models in reproducing observed light curves and
spectra in the optical and NIR.  The model {\sl 5p028z22.25} was
calculated previously with parameters appropriate for the class of
normal-bright SNe Ia \cite{PAH02}.
The two significant parameters affecting the chemical composition of
the model are central density ($\rho_c$) and the deflagration to
detonation transition density ($\rho_{tr}$).

The model has not been altered or tuned for this paper. It is based on
the explosion of a Chandrasekhar mass WD with a central density
$\rho_c = 2.0\times 10^9$ g cm$^{-3}$.  The WD evolved from a main
sequence progenitor of $5.0 M_{\odot}$ with solar metallicity
\cite{dominguez00}. The rate of burning during the deflagration phase
has been adjusted in order to reproduce 3-D results for the
deflagration. The transition from deflagration to detonation is
triggered when the density at the burning front drops below
$\rho_{tr}=2.5 \times 10^7$ g cm$^{-3}$.  From the physical point of
view, $\rho_{tr}$ should be regarded as a convenient way to adjust the
amount of material burned during the deflagration phase.

As mentioned above, the spectra provide direct information on the
velocity but not on the mass because only a fraction of the total mass
is involved in the spectral formation at a given time (H\"oflich
1995). 
We use a different approach to estimate the mass/velocity
relationship.  In Fig.~\ref{mass_vel}, the density and velocity
profiles are given as a function of the total mass. We use $v(M)$ to
estimate the amount of mass above the photosphere in the discussions
below. The reference model can be used for this estimate because, for
all DD models the entire WD is burned.  As a consequence, the values of
$v(M)$ and $\rho(M)$ as plotted in Fig.~\ref{mass_vel} are not
sensitive to specific model parameters (H\"oflich et al. 2002).

The chemical structure of our reference model is given in
Fig.~\ref{logyi}.  Qualitatively, the final burning products can be
understood in terms of the relation between the hydrodynamical and the
individual nuclear time scales.  The hydrodynamical time scale is
given by the total energy release during the explosion.    The nuclear
time scales are determined by the peak temperature during burning,
which depends on the energy release per volume because the energy
density is radiation dominated.  Thus, the density (and the  initial
C/O ratio) are  the dominant factors that determine the final
composition of a zone.  The actual density under which burning occurs
depends on the expansion of the WD during the explosion.  Therefore,
the distribution of the  nuclear burning products is  very model
dependent.  NIR spectra
reveal the chemical distribution of certain explosion products that
permit us to probe the explosion physics, in particular the
pre-expansion of the WD and progenitor properties.

In model regions of high density and, thus, high temperature ($T > 5
\times 10^9K$), burning proceeds up to nuclear statistical equilibrium
(NSE) and Fe-peak elements are produced.  The isotopic composition of
Fe-peak elements in the ejecta is determined by the electron captures
occurring at high temperatures and  densities near the center of the
white dwarf. As a result, the isotopic composition is highly sensitive
to the initial density of the dwarf and to the velocity of the
deflagration.  In our model, the region of complete burning to NSE
extends up to about 10,000 \kms.  Between $v_{exp} \approx
10,000-15,000$  \kms, intermediate mass elements are produced during
explosive oxygen burning.  A wide range of burning conditions ($ 5
\times 10^9K > T > 3 \times 10^9$) produces very similar element
ratios in the main products of burning (Si, S, and Ca).  Among the
products in this region of incomplete Si burning, only manganese and
vanadium exhibit a significant gradient in the quantity of product as
a function of velocity (or temperature). This makes Mn and V
potentially important diagnostic tools (see below).  The extreme outer
region ($T < 3 \times 10^9 K$ and $v_{exp} \gtrsim 15,000$ \kms)
undergoes explosive carbon burning with O, Mg and Ne being the main
products.

Detailed NLTE spectra and light curves have been calculated, based on
the explosion model.
The model computes the position of the photosphere and the line
forming region and calculates the relative line strengths.  Synthetic
spectra resulting from these calculations are shown in
Fig.~\ref{syn_spec}.  The model reaches a maximum brightness in B and
V of $-19.24$ and $-19.21$ magnitudes at about 17.5 and 18.5 days,
respectively.

The NIR is essential for measurement of freshly synthesized magnesium
because the high cross section of Mg lines in the optical and UV means
that very small amounts of Mg are required to form strong features in
those regions of the spectrum.   Magnesium lines in the optical and UV
will be dominated by absorption from primordial magnesium if the
progenitor has solar metallicity.  In a normal-bright SNe Ia up to
about maximum light, the photosphere is formed entirely in layers of
incomplete burning.  The spectra are dominated by elements of
incomplete burning such as Mg, Si, S and Ca. To form strong \mg\ lines
at $\approx 0.9$ \mum\ and $\approx 1.05$ \mum, the Mg abundance must
be larger than 1 to 2\% which is much greater than the primordial Mg
level.  Such a Mg abundance is expected from explosive carbon burning.
NIR lines from \mg\ are thus able to probe the region of transition
from carbon to oxygen burning but optical \mg\ lines do not.

Soon after maximum light, features from iron group elements begin to
dominate the spectrum.  Most prominent is the development of the
``pseudo'' emission features between 1.5 and 1.8 \mum\ which are
produced by blends of Co, Ni, and Fe lines.  Fig.~\ref{syn_spec}
illustrates the predicted development of this feature.  Line
blanketing increases the total opacity so that the features are formed
at a larger radius than the photosphere. In addition, the NIR lines
are due to transitions between highly excited atomic levels.  The
energy differences are small compared to the excitation energy, but
comparable to thermal energies.  The departure coefficients of the
upper and lower level are similar, and the source functions are
approximately a black body.  As a  result of the increased effective
area and high emissivity, the observed flux at these wavelengths is
increased.  The region from $1.1-1.5$ \mum\  has fewer of these iron
group lines and, consequently, the flux is lower.  Two peaks appear at
$\approx$ 1.60 \mum\ and 1.75 \mum\ with a notch between them that is
due to a gap in the blended lines at about $1.65-1.70$ \mum.   As
slower material passes through the photosphere, a broader range of
velocities is detected.  The features begin to smear out, and the gap
is gradually filled. Simultaneously, strong lines of Fe/Co/Ni appear
above 1.9 \mum\, and also at shorter wavelengths.   For more details,
see \cite{Wheeler98, PAH02}, and below.

\subsection{Doppler Shifts of Features}

During the first few weeks after the explosion, the primary opacity
source for SNe Ia in the NIR is Thomson scattering with some
contribution from free-free emission.  The continuum-forming region is
better defined in velocity in the NIR than it is in the optical and UV,
where resonance transitions influence the opacity and distort the
photon forming region.

In rapidly expanding, scattering dominated  envelopes, the lines are
P-Cygni like.  The absorption component of the line is formed when
photons from the photosphere are scattered out of the line of sight of
the observer into other directions and frequency bands.  
Photons may also be absorbed by an atom. In either case,
the spectrum will show an absorption feature, and  we can use the
Doppler shift of the absorption components of lines to measure the
expansion velocity of the line forming region.  In practice, the
radial change of the line source functions and opacities with distance
may introduce an error. From the  models, the  error in velocity is
estimated to be $\pm 5$ to $7 \%$ ($\approx 500$ \kms).

As expansion reduces the density and opacity, deeper layers are
revealed at later times, and the measured Doppler shift
decreases. However, once the layer containing a particular element has
completely passed outward through the photosphere, the observed
absorption feature from that element is characterized by a velocity
close to the inner edge of the layer. The feature will subsequently
decrease in strength as the column density drops due to the expansion
of the layer, but the Doppler shift of the absorption minimum will not
change.  In reality, lines at similar wavelength may complicate the
situation.  For instance, the \mg\ line at about 1.05 \mum\ has been
observed to show exactly this behavior of weakening while displaying a
constant Doppler shift (Bowers et al. 1998).  A \ion{Co}{2} line at a
similar wavelength occurs a few days after maximum light when the
photosphere is formed in the Fe/Ni core.  The presence of the
\ion{Co}{2} line will displace the apparent location of the absorption
minimum of \mg.

\section{Analysis of the Data}
\label{spectra}

As discussed in Section~\ref{datanal} we have identified the important
absorption lines and P-Cygni features in the spectra and measured the
observed expansion velocities.  The complete sample of thirteen
spectra is found in Fig.~\ref{sn13}.  The spectra are divided into two
groups for line identification and analysis.
The first group displays features from the outer layers of the
supernova before iron-group elements begin to dominate the spectra
(Fig.~\ref{e7}).  This group includes seven spectra obtained  between
-13 and +1 days of maximum light in the V-band ($V_{max}$).    The
second group contains six spectra obtained between +4 and +18 days
after $V_{max}$.  This group reveals the development of features
resulting from line blanketing due to thousands of iron-group lines
(Fig.~\ref{l6}).

At wavelengths between 1.31-1.38 and 1.78-1.88 \mum\ atmospheric
transmission is nearly zero ($< 10$\%) which results in very low
signal to noise ratios in these regions.  We have removed these
sections from the spectra obtained in cross dispersed mode.   In
addition, the spectra are particularly noisy at certain other
wavelengths due to low atmospheric transmission (50-70\%).  These
wavelengths are 0.93-0.96, 1.12-1.16 and 2.40-2.50 \mum\ in the
observed frame (nearly zero transmission from 2.5-3.5 \mum). The
locations of these regions in the rest frame of the host galaxy are
dependent on the redshift corrections.   For the spectra in our
sample, the regions of increased noise correspond to wavelength bands
from 0.90-0.94 \mum, 1.08-1.13 \mum, and red-ward of 2.35 \mum.

The spectra have been smoothed to reduce the noise in order to
facilitate identification of the spectral features.  
The amount of smoothing required to achieve a useful spectrum varies
widely depending on the quality of the data.  In each case, we have
attempted to balance the amount of smoothing against the benefits of
maintaining resolution.   The resolution for each spectrum after
smoothing is displayed in \kms\ at the upper right and corner of the
figures after the name of each supernova.  

Type Ia spectra are characterized by the steadily
diminishing flux at longer wavelengths.  As a result, the S/N of the
spectrum becomes worse at longer wavelengths because the level of the
signal from the supernova is not as far above background noise levels.
The spectrum from SN 2001dl is particularly noisy but is useful for
presenting the overall energy distribution from a Type Ia event at
that epoch.  A few features can be identified near the blue end of
such high noise spectra, but we are not able to define the position of
any other features in these spectra with less uncertainty than the
predicted line widths.

We examine the spectra beginning at shorter wavelengths and moving to
longer wavelengths.  Uncertainties in measuring the positions of
features in the spectra depend on the noise and amount of
smoothing, but typically range from $\pm $500-1400 \kms.  
The measurement uncertainties are noted in Table~\ref{v_table}.

\subsection{Six spectra from thirteen days before to one day after maximum 
light}
\label{early}

Seven NIR spectra from SNe Ia are presented in Fig.~\ref{e7}.  The
spectra were obtained at approximately -13, -6, -4, -3, -2, -1, and +1
days from maximum light.  The wavelength region in Fig.~\ref{e7} is
reduced from that of the full spectrum in order to emphasize the region
containing most of the features under discussion.  References to the
model in this section are discussed in more detail in
Section~\ref{models}.

The strong absorption feature found at 0.815-0.830 \mum\ is due to the
\ca\ ``NIR triplet'' (0.850, 0.854, 0.866 \mum). This feature has been
observed previously in optical spectra with extended coverage to 0.9
\mum.  The region from 0.801-0.893 \mum\ contains many \ca\ lines that
contribute to the broad P-Cygni shape of the feature.  The blue side
of the absorption feature includes contributions from \mn\ and
\ion{Co}{2}, the strongest of which is \ion{Co}{2} at 0.841 \mum, but
the \ca\ feature is more than 2000 times stronger.  The red side of
the feature is also influenced by \mn\ and \ion{Co}{2} with the
strongest line \ion{Co}{2} at 0.881 \mum.  
The expansion velocity of \ca\ in this feature is difficult to
determine because extensive blending obscures the location of the
absorption minimum for the individual lines.  Also, the lines from the
Ca triplet are so strong that the wings of the emission component
distort the absorption trough.  Measured velocities for \ca\ from this
feature in this group of spectra are between 10,000 and 12,000 \kms\
with uncertainties $\pm 1000-2000$ \kms.  That is in agreement with
the model predictions for the period before maximum light for \ca\
expansion velocities of $10,400-12,300$ \kms.

The \ca\ lines in the NIR triplet are formed well above the
photosphere and, even with solar abundance, a strong line is formed.
Therefore, this feature is not a good tracer for the location of
nuclear burning products \cite{hkt98}.

The broad feature found in the spectra from 0.88-0.94 \mum\ is
produced by 
the \mg\ doublet (0.922, 0.924 \mum).  At 2 or 3 days after $V_{max}$,
lines from \mn\ (0.944 \mum) and \ion{Co}{2} (0.955 \mum) contribute
in this region  and eventually dominate the \mg.  This behavior can be
seen in the spectra from the second group in our sample (+4 to +18
days; Fig.~\ref{l6}).  The blue edge of the \mg\ feature is used to
measure the Doppler velocity and we compare it with the 0.922 \mum\
line.  We find expansion velocities from $10,400-15,050$ \kms\ for
\mg\ based on these measurements although the velocity from the
earliest spectrum (SN 2002fk at -13 days) appears to be anomalously
low.  These velocities are consistent with standard models for SNe Ia
\cite{hk96}.   As discussed in  \S~\ref{models}, \mg\ is produced in
the outer layers of the supernova by explosive carbon burning.  \mg\
is not expected to be observed later than 2 or 3 days after maximum
light because expansion will reduce the column depth to the point
where \mg\ is not detectable.  The region from 0.90-0.94 \mum\ is also
one of the spectral regions with additional noise in the spectra due
to increased atmospheric opacity.

\mg\ is also detected near 1.04 \mum\ in these spectra.   This feature
is due to the \mg\ triplet at 1.091, 1.092, 1.095 \mum.
The detection of two \mg\ features (0.922 doublet and 1.091 triplet)
makes this a very confident identification.  Soon after maximum light,
there is a significant contribution on the red side of the \mg\
triplet feature from \ion{Co}{2} at 1.091 \mum\ since the \ion{Co}{2}
is produced closer to the center of the explosion and has a lower
expansion velocity.  At four days after maximum,  the \ion{Co}{2} line
is dominant and the \mg\ blended.  This behavior can be seen in the
spectra from the second group in our sample (+4 to +18 days;
Fig.~\ref{l6}).  The expansion velocities of \mg\ are measured at the
blue edge of the absorption minimum and compared to the 1.091 \mum\
component of the triplet.  We measure expansion velocities from this
feature to be 11,000-15,300 \kms.  That velocity range is consistent
with the results from the \mg\ doublet (0.922, 0.924 \mum) (see
Table~\ref{v_table} and Fig.~\ref{v_plot}) as well as the  predictions
for DD modelss.   Again, the velocity from the earliest spectrum (SN
2002fk at -13 days) appears to be anomalously low.

The broad P-Cygni feature with absorption minimum near 1.21 \mum\ is a
Calcium feature 
that appears a few days before maximum light and is due to a blend of
strong \ca\ lines from 1.23-1.29 \mum. The strongest lines in this
group are found at 1.243, 1.252, 1.268, and 1.283 \mum, but
identification of a single line is not possible.  We fit a curve to
the spectrum in order to find the minimum of the feature and associate
that minimum with the 1.268 \mum\ line.  The uncertainties in this
identification are estimated to be $\pm$ 900-1800 \kms.  The measured
expansion velocities for this feature are 13,400 to 15,200 \kms\ which
is above the model predictions for the period before maximum light of
10400-12300 \kms.  A few days after maximum light, the photosphere has
receded well within the Ca-rich layers and the line becomes too weak
to be observed.

The absorption minimum near 1.60 \mum\ is part of a P-Cygni like
feature with an emission peak near 1.68 \mum.  
This feature is likely  due to a blend of \ion{Si}{2} (1.691 \mum) and
\mg\ (1.676 \mum\ and others) with the \ion{Si}{2} dominant. We
estimate the location of the absorption minimum by fitting a curve to
the data. The spectra from SN 2000dn (-6d) and SN 2001dl (-3d) do not
provide clear minima and are not included in the tables.  This
\ion{Si}{2} identification is consistent with models that predict a
feature to be present until a few days after maximum light, but it is
not a definitive identification.  After +5 or 6 days, line blanketing
from thousands of Fe/Co/Ni lines creates pseudo emission features that
peak near 1.50 and 1.77 \mum\ (See Figure~\ref{l6}).  The notch
between these two emission peaks occurs at approximately the same
wavelength as the Mg/Si absorption minimum, but the features are not
related.

Significant features are not predicted or observed in the early K-band
(1.8-2.4 \mum) spectra of SNe Ia.  There is slight evidence in some of
the
spectra near maximum light for broad absorption features near 2.05 and
2.25 \mum\ (See Figure~\ref{sn13}).  If real, these would be due to
\ion{Co}{2} with the strongest lines at 2.135, 2.221, 2.250, 2.361 and
2.460 \mum.

\subsection{Seven Spectra from four to eighteen days after  maximum light}
\label{late}

The later spectra from our sample were obtained at approximately +4,
+5, +6, +10, +10, and +18 days after $V_{max}$ (Fig.~\ref{l6}).

The \ca\ triplet (0.850, 0.854, 0.866 \mum) is present in the large
P-Cygni feature as it was in the earlier spectra.  The measured
expansion velocity for \ca\ from this feature remains about 10,000
\kms\ although the photosphere has receded toward the center of the
explosion and much slower explosion products have been revealed (see
Table~\ref{v_table}.)  This is a consequence of the high cross section
of the \ca\ triplet which results in this feature being formed well
within the Ca-rich envelope.

The absorption feature visible at 0.89 \mum\ in the two earliest
spectra from this group is a blend of \ion{Si}{2} and \ion{S}{2}.

The feature near 0.91 \mum\ is attributed to a blend of \mn\ lines
from 0.931-0.955 \mum\ with the strongest the 0.944 \mum\ line.  At 12
days past maximum, blended iron group lines dominate this region of
the spectrum and \mn\ is no longer detectable.  The \mn\ velocities
clearly diminish with time and trace the movement of the photosphere
through the ejecta (See Table~\ref{v_table} and Figure~\ref{v_plot}.)

The 0.91 \mum\ feature can clearly be seen in the spectra obtained
from four to six days after maximum light and possibly also at +10
days.  That is consistent with model predictions that the \mn\ line
will be weak or blended until 4 or 5 days after maximum light.
However, since all line identifications are model dependent and we do
not find a second \mn\ feature to corroborate line strengths and
velocities, we must consider the \mn\ identification to be uncertain.
The location of the \mn\ feature is also near the $0.90-0.94$ \mum\
band of increased noise due to atmospheric opacity.  Expansion
velocities measured for \mn\ are 11,100 \kms\ at +4 days and fall to
6,350 \kms\ at +10 days.  The observed velocities for \mn\ diminish
with time as the photosphere recedes towards the center of the
supernova (See Table~\ref{v_table}).

Mn is important because it is a product of incomplete silicon burning
during the detonation phase of the SNe Ia explosion.  \mn\ is a
sensitive diagnostic of explosion dynamics because the production of
\mn\ changes by two orders of magnitude depending on burning
temperature.  This is in contrast to other products of partial silicon
burning which display equal proportions throughout the range of
temperatures  (See Fig.~\ref{mass_vel} and Section~\ref{models}).

The 0.99 \mum\ feature, which first becomes visible four days after
maximum light and becomes more prominent through +12 days, is due to
\ion{Fe}{2}.
A \ion{Co}{2} feature from the 1.091 \mum\ line is apparent in the
spectra from this group.
The \ca\ feature with absorption minimum near 1.21 \mum\ is visible up
to about one week after maximum light.  That is consistent with model
predictions.  The measured velocities for \ca\ from this feature follow
the reduction in photospheric velocity with time (See
Table~\ref{v_table}).

The four latest spectra in our sample, obtained +6, +10, +10, and +18
past maximum light, exhibit a reduced continuum level from $1.1-1.5$
\mum\ as  compared with the earlier spectra.  Fig.~\ref{sn13} clearly
shows this effect.  This change in slope is due to the region having
fewer blends of iron group lines compared to adjacent wavelengths
\cite{Wheeler98}.  Note that the spectrum of SN 2001bf (+5d) does not
exhibit any deficit in this region, even though the epoch is estimated
to be only one day from that of SN 2000dk (+6d) (Fig.~\ref{l6}).

The emission peaks at  1.60, 1.75, 2.15, 2.25, and 2.35 \mum\ are the
result of increased flux due to line blending from iron group
elements.  The radius of optical depth unity is increased at these
wavelengths producing an increase in observed flux due to the greater
effective area of radiation (Compare Fig.~\ref{l6} and
Fig.~\ref{syn_spec}).

The abrupt edge on the blue side of the peak found near 1.55 \mum\
defines the transition from partial to complete Silicon burning.  Due
to the high optical depth, this feature is formed close to the outer
edge of the Fe/Co/Ni core. The rest  wavelength of  this feature  is
$\approx 1.57$ \mum.  The observed velocities are above 10,000 \kms\
for the spectra obtained from +4 to +10 days after maximum light and
9,800  \kms\ for the spectrum obtained at +18 days.  This result is
consistent with model predictions.  Fig.~\ref{logyi} shows all silicon
to be consumed below $\approx 8500$ \kms\ for a model of a
normal-bright Type Ia.

\section{Constraints on Supernovae Models}
\label{models}

\subsection{Outer layers}

{\bf Residual Carbon and He:}  There is no evidence in our spectra for
\ion{C}{1} at 0.93 or 1.13 \mum.  That implies that basically all
material in the WD has been burned.  If \ion{C}{1} were present, it
should be observable \cite{PAH02}.  
Fisher, et al. discuss possible
identifications of \ion{C}{2} at velocities $>$ 15,000 \kms in spectra
from SN 1990T and $>$ 25,000 \kms\ for SN 1990N \cite{fisher98}.  If
these identifications are correct, Figure~\ref{mass_vel} may be used
to  show that the mass fraction above 15,000 \kms\ is less than 10\%
of the WD progenitor mass and for 26,000 \kms, the mass of unburned
material is $\le$ 1\% of the progenitor mass.   Consequently, burning
must have reached the outer layers.

We also do not see any indication for \ion{He}{1} from the 2.058 \mum\
line. Combined with the detection of several \mg\ lines, the lack of
He supports the identification of the 1.05 \mum\ feature as \mg\ rather
than \ion{He}{1} at 1.083 \mum.

{\bf Magnesium II Velocities:} Mg is a product of explosive carbon
burning without oxygen burning. Due to the higher burning temperatures
during O burning, heavier elements are produced further along the
alpha chain (H\"oflich  1997, Wheeler, et al. 1998). \mg\ lines have
been observed at $\approx$ 0.9 and 1.05 \mum\ from pre-maximum to a
few days after maximum light as expected.
After 5 or 6  days before maximum light, the continuum forming region
will be well beneath the magnesium-rich region.  Consequently, the
Doppler shift of the absorption minimum will represent the inner edge
of that region.  We measure expansion velocities for \mg\ between
12,450 and 15,300 \kms.

Figure~\ref{mass_vel} can be used with the \mg\ velocities to put
limits on the amount of unburned matter after the explosion.
Depending on the supernova, less than $0.2 M_\odot$ (SN 2002hw, SN
2000dm) or $0.1 M_\odot$ (SN 2000en, SN 2001br) remain outside of the
region of carbon burning and could possibly be unburned.  These values
are upper limits because \mg\ is present at higher velocities in the
blue wings of the lines, meaning that even less material remains
unburned.  In principle, the limits on the mass of unburned material
can be improved by measuring the blue wings of the \mg\ features, but
the S/N ratio in our data severely limits this approach.  This result
for the amount of unburned matter is consistent with the limits due to
\ion{C}{1} as discussed above.

\subsection{Layers of Silicon Burning}

Silicon continues to burn under conditions where both carbon and
oxygen are completely consumed.  The products of Si burning are
intermediate mass elements at burning temperatures less than
$\approx5\times10^9$ K.  Above that temperature, Si is burned entirely
to iron peak elements.  In mass space, the region of of incomplete Si
burning extends from $\approx$ 9,000 $-$ 16,000 \kms\ (see
Figure~\ref{logyi}).  The figure also shows that below $\approx$ 8,500
\kms\ Si is completely consumed.

{\bf Manganese Velocities:} Mn is a product of incomplete silicon
burning, but the quantity of Mn generated during the explosion depends
sensitively on burning temperatures.  The plot from the reference
model of mass fraction of explosion products as a function of
expansion velocity (Fig.~\ref{logyi}) reveals silicon (depicted as the
solid green line) at the top of the region from 9,000 $-$ 16,000 \kms\
in the reference model.  The figure also shows that this velocity
region contains sulfur, argon, calcium, manganese and vanadium.  Among
the products of incomplete Si burning, Mn exhibits the strongest
gradient in the quantity of product as a function of velocity.  It is
also produced in sufficient quantities to form detectable features.
Since the velocity of the ejecta is proportional to the radius from
the center of the explosion, manganese can also be used to measure the
temperature of the burning because temperatures are hotter nearer to
the core and more Mn is produced.

The observed velocity of \mn\ recedes with the photospheric velocity
from 11,000 \kms at +4 days to 6,350 \kms at +10 days (See
Table~\ref{v_table}).  The \mn\ feature is observed to become broader
at later epochs.  From the models, the increase in mass fraction for
\mn\ at the lower velocities (due to higher burning temperatures) is
consistent with the broader absorption features.

{\bf Silicon to Nickel Transition:}  The short wavelength edge near
1.55 \mum\ is a good measure of the size of the Fe/Co/Ni core
(Fig.~\ref{l6}). The spectra in our sample have transition velocities
of $\gtrsim 9,800-11,200$ \kms.  The Doppler shift is based on a rest
value for the wavelength of this transition of 1.570 \mum.  The
velocity at the edge recedes slightly, but remains above 9,800 \kms\
at 18 days after maximum light at which time the photospheric
velocity is down to about 3,000 \kms.  
Pseudo emission features like this are due to line blanketing that
increases the total opacity so that the features are formed at a
larger radius than the photosphere. As a result of the increased
effective area and high emissivity, the observed flux at these
wavelengths is increased (See Section~\ref{rmod}).  Notice the notch
at about 1.65-1.75 \mum\ which is well developed at +10 days in the
spectrum of SN 2001bg (Fig.~\ref{l6}). As predicted (See
Section~\ref{rmod}), the notch should become partially filled as time
advances. This effect can be seen in the spectral development between
SN 2001bg (+10d) and SN 2001en (+18d).  Note that the velocity  of the
partial to complete Si burning transition is very sensitive to the
transition density parameter ($\rho_{tr}$) in DD models \cite{h95,
hk96, PAH02}. Higher resolution spectra have the potential to
constrain this important parameter.

\section{Conclusions}
\label{conc}

We have studied a set of thirteen early time, near NIR-spectra (0.8 to
2.4 \mum) of ``Branch-normal'' Type Ia supernovae. The observations
more than double the number of SNe Ia in the literature for which NIR
spectra are available near maximum light.  The spectra were obtained
at the NASA Infrared Telescope Facility at resolutions between 250 and
1200 at an apparent brightness between +14 and +18 mag. NIR
observations are critical because certain elements have no lines at
optical wavelengths, are heavily blended in the optical region or are
so strong that even solar metallicities produce highly saturated
lines.  The wide wavelength range of our sample allows testing of line
identifications by the presence of lines of the same ion or element at
other wavelengths.  The low brightness limit of our program enabled us
to guarantee appropriate targets for observations scheduled months in
advance. We are unable to observe individual SNe Ia in a time sequence
but the measurements in our sample are spread over a range of epochs.
The calibration of the flux distribution has been tested based on both
broad band filters and background subtraction using the data.  The
methods are consistent within $\approx 20\%$.  The quality of the
spectra is sufficient to put strong constraints on models for SNe~Ia.

As a reference, and to identify the spectral features, we have chosen
a delayed detonation model for Branch-normal SNe Ia. The use of a
particular model and explosion scenario does not a priori exclude
other scenarios. Pure deflagration models and mergers are similar in
the element production.  However, edge-lit Helium detonation models
show Ni and He at high velocities which should be observable, but are
not found in the spectra from our sample.  The various explosion
scenarios and individual realizations due to variation in model
parameters can be
differentiated by the separation of spatial element distributions of
several thousand \kms.  Consequently, the resolution of our spectra is
sufficient to distinguish between models.

To analyze the expansion velocity of individual SNe Ia, we used the
Doppler shifts of the absorption component. Based on the model, this
approach yields an internal accuracy of about 5 to 7 \% which is
consistent with the resolution of the observations.  The uncertainties
from the measurements are dominated by the S/N ratio and are of the
order of $\pm$ 1,000 to 1,500 \kms .

All strong features in the spectra could be identified in the entire
set of our observations. They are consistent with the reference model.
If ordered as a time sequence, the expansion velocities recede with
time as expected for the homogeneous group of Branch-normal SNe Ia. We
find no correlation between the $\Delta m_{15}$ parameter and
the expansion velocity of explosion products.

No evidence for the \ion{He}{1} line at $\approx 2.05$ \mum\ was
found.  He is the characteristic feature for sub-Chandrasekhar mass
WDs which are edge-lit by a He layer (Nomoto et al. 1980, Woosley \&
Weaver, 1994).  

One of the main results is that the overall chemistry shows a radially
layered structure for the matter which has undergone explosive carbon,
incomplete and complete Si burning.  In our reference model, the
photospheric velocity is at $\approx 12,000$ \kms\ three days before
maximum light and 9,500 \kms\ at five days after maximum.
Based on the observations between -6 and +1 days, \mg\ expansion
velocities are between $\approx$ 12,000 and $\approx$ 15,000 \kms.
These values are consistent with values observed in 1994D (14,000
\kms, Wheeler et al. 1998) and SN 2000el (18,000 \kms, Rudy et
al. 2002).  (Note that our velocities are based on the measurement of
two \mg\ lines which increases the level of confidence in the
results.)  We conclude that the photosphere must be inside and well
separated from the entire \mg\ envelope.  That means that we are
observing the low velocity edge for \mg\ in each supernova.
Consequently, there is a well defined transition from layers that
experienced explosive C burning which produces \mg\ to layers with
both C and O burning as predicted by the model (Fig.~\ref{logyi}).

Similarly, the Fe II emission-feature at $\approx$ 1.5 to 1.9 \mum\
indicates a well-defined transition from incomplete Si burning to
burning up to NSE at $ \gtrsim 10,000$ \kms.  This result is
consistent with optical spectra which typically give lower limits for
the \ion{Si}{2} velocity $\geq 9,000 - 10,000$ \kms (Barbon et
al. 1990).

The presence of \mg\ in the NIR provides a limit for the amount of
unburned material in the outer layers of $\leq 0.1 M_\odot$. This is
an upper limit because if Mg is present further out, then the amount
of unburned material will be even less.  The use of the blue edge of
the \mg\ may provide better limits, but the S/N in our data does not
allow measurement of the extension in the line wing. 

In addition, we do not see evidence for large scale mixing. Iron peak
elements are not observed at high velocities and intermediate mass
elements are not observed  at low velocities.   No unburned carbon is
found in our data.

Distinct chemical layering, as suggested by our observations, is a
signature of a detonation. Combined with the lack of carbon, our
results strongly favor the presence of a detonation phase during the
explosion.  As discussed in the introduction, deflagration fronts seem
to produce structures that are dominated by large scale mixing.  They
include unburned or partially burned material close to the center.
Current 2-D \cite{livne93, lisewski00, reinecke99} and 3D calculations
\cite{khokhlov95, khokhlov01} of pure deflagration models show that $>
0.4 M_\odot$ remain unburned.  The existence of a layered structure and
the limit of a very small amount of unburned matter are inconsistent
with a pure deflagration and suggest the presence of a detonation.

For merger scenarios, we expect the burning front to propagate as a
detonation (Benz et al. 1992, Khokhlov et al. 1994, H\"oflich \&
Khokhlov 1996).  That would produce a layered chemical structure, but
a significant amount of unburned, C/O rich, matter is predicted to
remain in the outer layers and should be observed during the early
times (Khokhlov et al. 1994, H\"oflich \& Khokhlov 1996, H\"oflich et
al. 2002).  Again, the lack of \ion{C}{1} in our seven earliest
spectra is inconsistent with merger models.

Overall, the supernovae in our sample are rather homogeneous; however,
measured expansion velocities show variations between SNe at a given
epoch of a few thousand \kms\ (see Figure~\ref{sne} and
Table~\ref{v_table}).  This clearly points towards variations among
the group of Branch-normal SNe Ia.  We find no correlation between
expansion velocity and the decline rate parameter, $\Delta m_{15}$.
Since NIR lines contain independent information, the properties of NIR
lines may be used to get a handle on the variety in the observations
and to provide the refinements needed to improve the accuracy of
relative distances determined by SNe Ia beyond the level of the
brightness decline relation.  Given the lack of calibrated LCs, a
detailed evaluation is beyond the scope of this paper.

The possible identification of a \mn\ line at $\approx$ 0.94 \mum\ in
the spectra obtained after +3 days post maximum is an important
result.  This identification is consistent with the theoretical model
prediction both with respect to the time it appears and the Doppler
shifts.  \mn\ promises to be a valuable diagnostic as more
measurements are obtained.  It is a unique probe of the region of
incomplete silicon burning due to the gradient of \mn\ production
through this region while the level of other burning products remains
nearly constant.

In conclusion, the results of our analysis are consistent with delayed
detonation (DD) models of SNe Ia.  The data strongly favor models
where the entire progenitor, including the outermost layers, undergo
burning and that there is no strong mixing of the chemical layers.

Finally, we mention the limitations of this work which should be seen
as a step toward further and more detailed studies.  Detailed optical
spectra and calibrated light curves were not available for this study.
Consequently, we are unable to test whether or not it is possible to
explain or compensate for the variations in our sample using
variations in peak brightness.  For most of the objects, corresponding
data have been obtained elsewhere and will be available at a later
time.  After testing maximum brightness data, the variations will be
analyzed using changes in the model parameters that define the
progenitor WD, such as central density, mass on the main sequence and
metallicity.

Although our measurements represent a significant increase of the
number of the available NIR spectra, we are still dealing with small
number statistics. Better statistics are needed to understand the
variations within our sample. Moreover, merging WD are apparently not
found in our sample, but this class of objects may contribute to the
SNe Ia population in a significant number.  Higher signal to noise
ratios are needed to accurately measure the extension of the blue
wings of
\mg\ to improve the limits on the unburned, outer layers, and to
increase the accuracy of velocity measurements.  Higher S/N would
provide tighter limits on the possible mixing of different elements,
and detailed fits to individual observations should be performed,
including detailed 3-D models.

\noindent{\bf Acknowledgments:}

We thank W. D. Li and M. Papenkova for providing light curve data for
many of the SNe in our sample.  We thank the individuals at the IRTF
for guidance and help with the observations.  In particular, Alan
Tokanaga, Bobby Bus, John Rayner, Mike Cushing, Karen Hughes, Tony
Denault, Bill Golisch,  Dave Griep, and Paul Sears have been most
helpful.  We would also like to thank the TAC of the IRTF for support
and instructive comments.  GHM would like to thank Dan Jaffe and
Weidong Li for support and helpful comments.  This research has made
use of the NASA/IPAC Extragalactic Database (NED) which is operated by
the Jet Propulsion Laboratory, California Institute of Technology,
under contract with the National Aeronautics and Space Administration.
This research is supported in part by NSF grant 0098644 and by NASA
grant NAG5-7937.

\appendix
\section {The Supernovae (in order of discovery)}
\label{sn}

Observational details are compiled in Table~\ref{obs_table} and
Table~\ref{norm_table}.  Additional information on the techniques used
to establish the epoch of observation and $M_V$ are found in
\S~\ref{sne}.  We attempt to establish the date and apparent maximum
brightness in the V-band for each SNe in our sample.
Where we have good B-band data, the date of $V_{max}$ is taken to be
the date of $B_{max} +2$ days.  The $\Delta m_{15}$ values are
measured directly from the photometric data and are not calculated by
fitting to template light curves.


\subsection {SN 2000dk}
SN 2000dk is the only event in our sample to be hosted by an
elliptical galaxy.  The SN was discovered in NGC 382 on Sept. 18, 2000
\cite{Beckmann_7493}.  The recession velocity for this galaxy is 5228
\kms.  At discovery, the V-band magnitude of the object was 16.0.  A
spectrum taken Sept. 20 (JD 2451808) by the CfA group \cite{Jha_7494},
established SN 2000dk as a Type Ia and estimated the epoch to be $-2
\pm 3$d which corresponds to a maximum light date of $2451810 \pm 3$d.
Photometry from the CfA group indicates a $V_{max}$ date of 2451812.5
\cite{Jha02}.  A relative B-band light curve for SN 2000dk finds
$B_{max}=2451811.8 \pm 1$d.  AUDE suggests that the maximum light was
$2451811 \pm 3$d at $V\sim 15.2 \pm  0.2$.  The VSnet light curve
suggests a maximum light date was $245810 \pm 4$d at $V\sim 15.2 \pm
0.5$.

We estimate for SN 2000dk that maximum V light occurred JD $2451814
\pm3$d, at apparent $V_{mag} = 15.2 \pm 0.5$.  The decline parameter
$\Delta m_{15} = 1.5 \pm0.2$ mag.

\subsection {SN 2000dm}
SN 2000dm was discovered in the spiral (Sab) galaxy UGC 11198 on
Sept. 24, 2000 by Aazami and Li \cite{Aazami_7495} at $V\sim16.1$.
The recession velocity for this galaxy is 4507 \kms.  The supernova
type was determined by the CfA group from a spectrum obtained
Sept. 25.1 (JD 2451812.5) \cite{Jha_7497a}.  The estimated epoch of
$-1 \pm 2$d, corresponds to a maximum light date of 2451813.5.  In the
same circular, Fillipenko and Chornock \cite{Filippenko_7497b} confirm
the supernova type and define the epoch as very close to maximum
brightness.  The relative B-band light curve for SN 2000dm indicates
that $B_{max}$ occurred between JD 2451815.6 and 2451817.6.  The VSnet
light curve shows SN 2000dm becoming brighter up to Sept. 30.8 (JD
2451817.6) at $V\sim 15.2$, but there is a subsequent gap in the data
until Oct. 08 at $V\sim 15.5$.  The AUDE data reveal increasing
brightness up to JD 2451817.3 at $V\sim15.1$.

We estimate for SN 2000dm that maximum V light occurred JD $2451818
\pm3$d, at apparent $V_{mag} = 15.1 \pm 0.5$.  The decline parameter
$\Delta m_{15} = 1.2 \pm0.2$ mag.

\subsection {SN 2000dn}
SN 2000dn is the most distant SN in our sample with a recession
velocity for the host galaxy of 9613 \kms.  It was discovered in the
spiral ((R')SAB0+ pec) galaxy IC 1468 on Sept. 27, 2000 \cite{Yu_7498}
at $V\sim17.9$.  A CfA spectrum obtained Sept. 29.3 (JD 2451816.8)
\cite{Jha_7499} established the type and estimated the epoch to be $-3
\pm 2$d, corresponding to a maximum light date of JD 2451819.8.  The
VSnet light curve agrees with a maximum light date of JD 2451820 $\pm
5$ at $V\sim 17.0 \pm 0.4$.  A relative B-band light curve for SN
2000dn indicates that  $B_{max}=2451823.7 \pm 2$d.

We estimate for SN 2000dn that maximum V light occurred JD $2451826
\pm3$d, at apparent $V_{mag} = 17.0 \pm 0.5$.  The decline parameter
$\Delta m_{15} = 0.8 \pm0.2$ mag.

\subsection {SN 2000do}
SN 2000do was discovered in the spiral (Sc) galaxy NGC 6754 on
Sept. 30, 2000 \cite{White_7500a}. The recession velocity for this
galaxy is 3257 \kms.  At discovery $V\sim15.6$.  Suntzeff
\cite{Suntzeff_7500b} established the type from a spectrum obtained
Sept. 30 and estimated the epoch to be a few days after maximum.
Unfiltered magnitude estimates by Bembrick \cite{Bembric_7509} show
the object becoming dimmer beginning at $V=15.0$ on Sept. 30 (JD
2451818.5).  The maximum light date must therefore precede discovery.
There are no light curves available for this object.

We estimate for SN 2000do that maximum V light occurred JD $2451816
\pm5$d, at apparent $V_{mag} = 14.6 \pm 0.5$.

\subsection {SN 2001bf} 
The unclassified galaxy MCG +04-42-022 was host to SN 2001bf which was
discovered on May 3, 2001 \cite{Hurst_7620} at $V\sim16.5$.  The
recession velocity for the host is 4647 \kms.  The supernova type was
determined by the GAO group from a spectrum obtained May 11 (JD
2452040.8) \cite{Kawakita_7625a}. The epoch is estimated to be shortly
before maximum light.  Unfiltered magnitude estimates by West
\cite{West_7625b} show the object becoming brighter from May 7 to May
8 and reaching $R=14.5 \pm0.2$ on May 14.4 (JD 2452043.9).  Although
Phillips and Krisciunas tentatively identified SN 2001bf as a Type Ic
\cite{Phillips_7636}, Chornock, Fillippenko and Li \cite{Chornoc_7701}
confidently reconfirmed the original designation of the object as a
Type Ia.  Their spectrum, obtained May 17 (JD 2452046.8), affirms the
supernova type and the epoch is estimated to be near maximum light.
VSnet data suggest a maximum light date of JD $2452045 \pm 5$ at
$V\sim 14.7^{+0.5}_{-0.2}$.  The AUDE data provide JD $2452044.6 \pm
5$ as the maximum light date at $V\sim 14.7^{+0.5}_{-0.2}$.  A
relative B-band light curve for SN 2001bf gives $B_{max}=2452043.9 \pm
2$d.  Given that the $R=14.5\pm0.2$ measurement corresponds to
$V\sim14.8\pm0.2$ and the scatter in the amateur data favor a small
reduction in maximum brightness for SN 2001bf, we estimate
$V_{max}=14.8$.

We estimate for SN 2001bf that maximum V light occurred JD $2452046
\pm3$d, at apparent $V_{mag} = 14.8 \pm 0.5$.  The decline parameter $\Delta m_{15} = 0.8
\pm0.2$ mag.

\subsection {SN 2001bg}
SN 2001bg was discovered at $V\sim14.0$ in the spiral (SBb/Sc) galaxy
NGC 2608 on May 8, 2001 \cite{Hurst_7621}.   With a recession velocity
for the host of 2135 \kms \cite{NED} this is the closest SN in our
sample.   Separate determinations of the supernova type were made by
the Tel Aviv University group \cite{Gal-Yam_7622a} and GAO group
\cite{Kawakita_7622b} from spectra obtained May 10 (JD 2452040). The
epoch is estimated to be near or shortly before maximum light.   A CfA
spectrum obtained May 15 (JD 2452045) \cite{Matheson_7626} confirms SN
2001bg to be a Type Ia supernova and the epoch is estimated to be $5
\pm 2$ days after maximum light corresponding to a maximum light date
of JD 2452040.  CCD R-band photometry by Horoch \cite{Hornoch_7639}
provides a maximum light date of JD 2452041.  The relative B-band
light curve for SN 2001bg finds $B_{max}=2452040.7 \pm 1$d. The VSnet
data suggest a maximum light date of JD $2452043 \pm 5$ at $V\sim 13.7
\pm 0.5$.  AUDE data imply a JD $2452042 \pm 4$ maximum light date at
$V\sim 13.6 \pm 0.3$.

We estimate for SN 2001bg that maximum V light occurred JD $2452043
\pm3$d, at apparent $V_{mag} = 13.7 \pm 0.5$.  The decline parameter
$\Delta m_{15} = 1.0 \pm0.2$ mag.

\subsection {SN 2001br}
The host for SN 2001br is the spiral (SBa) galaxy UGC 11260.  The SN
was discovered on May 13, 2001 \cite{Hurst_7629a} at $V\sim16.6$.  The
recession velocity for this galaxy is 6184 \kms.  The discovery
announcement also reported the object to be brightening from $V\sim
16.6$ on May 13 to $V\sim 16.0$ on May 21 (JD 2452051).  The supernova
type was established by the CfA group from a spectrum obtained May 21
(JD 2452051) \cite{Matheson_7629b}. The epoch was estimated to be $-1
\pm 2$d, corresponding to a maximum light date of JD 2452052.  No
amateur data is recorded for this SN.  The relative B-band light curve
for SN 2001br finds $B_{max}=2452053.0 \pm 1$d.

We estimate for SN 2001br that maximum V light occurred JD $2452055
\pm3$d, at apparent $V_{mag} = 16.0 \pm 0.5$.  The decline parameter
$\Delta m_{15} = 0.9 \pm0.2$ mag.

\subsection {SN 2001dl}
SN 2001dl was discovered in the spiral (Sc/I) galaxy UGC 11725 on July
30, 2001 \cite{Yu_7675}.  The recession velocity for this galaxy is
6204 \kms.  At discovery $V\sim17.1$.  A spectrum obtained August 8
(JD 2452130) \cite{Patat_7680} determined the supernova type and the
epoch was estimated to be at maximum light. The VSnet data suggest a
maximum light date of August 7 (JD $2452129 \pm 5$d) at $V\sim 16.0
\pm 0.5$.   AUDE data imply a Aug. 7 JD $2452129\pm 5$ maximum light
date at $V\sim 16.0 \pm 0.4$. A relative B-band light curve for SN
2001dl indicates that $B_{max}=2452130.9 \pm 1$d.

We estimate for SN 2001dl that maximum V light occurred JD $2452133
\pm3$d, at apparent $V_{mag} = 16.0 \pm 0.5$.  The decline parameter
$\Delta m_{15} = 0.8 \pm0.2$ mag.

\subsection {SN 2001en}
SN 2001en was discovered in the peculiar, triple cored galaxy NGC 523
at $V\sim17.5$ on Sept. 26, 2001 by Hutchings and Li
\cite{Hutching_7724} and also on Sept. 27 by Zhou and Li
\cite{Zhou_7725}.  The recession velocity for the host galaxy is 4758
\kms.  The supernova type was established from a NIR spectrum obtained
Oct. 8 by \cite{Marion_7732a}. They estimated the epoch to be near
maximum light. The type was confirmed by an optical spectrum obtained
Oct. 11 (JD 2452194) by Matheson et al. (2001c). That group estimated
the  epoch to be at maximum light $0 \pm 3$. A V-band light curve by
The Lick Observatory and Tenegra Observatory Supernova Searches
\cite{LOTOSS} suggests a maximum light date of JD $2452194 \pm 2$ at
$V\sim 15.0 \pm 0.1$. The maximum light date from the VSnet data is JD
$2452195 \pm 4$ at $V\sim 14.5 \pm 0.5$.  The AUDE data suggest JD
$2452196 \pm 4$ for the maximum light date at $V\sim 14.5 \pm 0.4$.  A
relative B-band light curve for SN 2001dl finds $B_{max}=2452192.8 \pm
1$d.

We estimate for SN 2001en that maximum V light occurred JD $2452195
\pm3$d, at apparent $V_{mag} = 15.0 \pm 0.2$.  The decline parameter
$\Delta m_{15} = 1.1 \pm0.2$ mag.


\subsection {SN 2002fk}
SN 2002fk was discovered in the spiral galaxy (SA(s)bc) NGC 1309 at
$V\sim15.0$ on Sept. 17.7, 2002 by R. Kushida \cite{kushida_7973} and
also by J. Wang and Y. L. Qiu \cite{wang_7973}.  The recession
velocity for the host galaxy is 2135 \kms.  The supernova type was
established by an optical spectrum obtained Sept. 20 (JD 2452538)
\cite{aynani_7976}. The epoch was estimated to be before  maximum
light.  The maximum light date from the VSnet data is JD $245249 \pm
4$ at $V\sim 13.2 \pm 0.5$.  A relative B-band light curve for SN
2002fk finds $B_{max}=2452548.8 \pm 1$d.

We estimate for SN 2002fk that maximum V light occurred JD $2452550
\pm3$d, at apparent $V_{mag} = 13.2 \pm 0.2$.  The decline parameter
$\Delta m_{15} = 1.1 \pm0.2$ mag.

\subsection {SN 2002ha}
SN 2002ha was discovered in the spiral galaxy (SAB(r)ab) NGC 6962 at
$V\sim17.3$ on Oct. 21.2, 2002 by Graham, Panankova and Li
\cite{graham_7997}.  The recession velocity for the host galaxy is
4211 \kms.  The supernova type was established by an optical spectrum
obtained Oct. 25.0 (JD 2452572.5) \cite{hamuy_7999}. The epoch was
estimated to be three days before maximum light.  The maximum light
date from the VSnet data is JD $2452583 \pm 4$ at $V\sim 14.7 \pm
0.5$.  The AUDE data suggest JD $2452582 \pm 4$ for the maximum light
date at $V\sim 14.5 \pm 0.5$.

We estimate for SN 2002ha that maximum V light occurred JD $2452583
\pm4$d, at apparent $V_{mag} = 14.7 \pm 0.5$.

\subsection {SN 2002hw}
SN 2002hw was discovered in the spiral galaxy (SAc) UGC 52 at
$V\sim16.8$ on Nov. 9.2, 2002 by Schwartz and Li \cite{schwartz_8014}.
The recession velocity for the host galaxy is 5257 \kms.  The
supernova type was established by an optical spectrum obtained
Nov. 11.0 (JD 2452589.5) by three teams cited in IAUC 8015 (Hamuy
2002, Filippenko \& Chornock 2002, Matheson, Challis and Kirshner
2002). The epoch was estimated by Hamuy and Filipenko and Chornock to
be near maximum light.  Matheson, Challis and Kirshner estimated the
epoch as before maximum and suggested that the slope of the continuum
indicated significant extinction due to dust.  The maximum light date
from both the VSnet and AUDE data is between JD 2452590 and 2452605 at
$V\sim 16.2 \pm 0.5$.

We estimate for SN 2002hw that maximum V light occurred JD $2452594
\pm4$d, at apparent $V_{mag} = 16.2 \pm 0.5$.



\begin{deluxetable}{llcrrrrrr} 

\tabletypesize{\scriptsize} \tablecolumns{9}  \tablewidth{0pc}  
\tablecaption{Observational and Spectral Parameters \label{obs_table}}  
\tablehead{\colhead{SN} &  \colhead{Obs. Date} &  \colhead{JD} &  \colhead{Max. Light\tablenotemark{a}} & 
\colhead{Epoch (V)\tablenotemark{b}} & \colhead{Est. $V_{obs}$\tablenotemark{c}} &
\colhead{Exp. Time\tablenotemark{d}} &  \colhead{Inst. Res.} & \colhead{S/N\tablenotemark{e}}\\ 
\colhead{} & \colhead{(UT)} & \colhead{(2450000+)} & \colhead{(2450000+)}  & \colhead{(days)} & 
\colhead{(mag)} &\colhead{(mins.)} & \colhead{($d\lambda/\lambda$)} & \colhead{} }

\startdata 
2000dk    & Oct. 2.5  & 1820.0  & $B=1811.9 \pm2$      & $+6\pm3$  & 16.0 & 50  & 1200 & 16\\
2000dm    & Oct. 1.3  & 1818.8  & $B=1815.6 \pm2$      & $+1\pm3$  & 16.1 & 50  & 750  & 21\\
2000dn    & Oct. 2.4  & 1819.9  & $B=1823.7 \pm2$      & $-6\pm3$  & 17.9 & 25  & 250  & 32\tablenotemark{f} \\
2000do    & Oct. 2.2  & 1819.7  & $V=1816   \pm5$      & $+4\pm5$  & 15.6 & 25  & 750  & 11\\
2001bf    & May 22.4  & 2050.9  & $B=2043.9 \pm2$      & $+5\pm3$  & 16.0 & 150 & 750  & 34\\
2001bg    & May 23.3  & 2052.8  & $B=2040.7 \pm1$      & $+10\pm2$ & 14.0 & 50  & 1200 & 18\\
2001br    & May 23.7  & 2053.1  & $B=2053.0 \pm1$      & $-2\pm2$  & 16.6 & 100 & 750  & 16 \\
2001dl    & Aug. 12.4 & 2129.9  & $B=2130.9 \pm2$      & $-3\pm3$  & 17.4 & 25  & 750  & 6\\
2001en\_1 & Oct. 8.3  & 2190.8  & $B=2192.8 \pm1$      & $-4\pm2$  & 14.5 & 50  & 1200 & 14 \\
2001en\_2 & Oct. 30.3 & 2212.8  & $B=2192.8 \pm1$      & $+18\pm2$ & 17.2 & 50  & 250  & 36\tablenotemark{f}\\
2002fk    & Sept. 19.5& 2537.0  & $B=2548.8 \pm1$     & $-13\pm2$ & 14.0 & 125 & 1200 & 36\\
2002ha    & Nov. 14.3 & 2592.7  & $V=2583   \pm4$      & $+10\pm4$ & 15.5 & 25  & 250  & 36\tablenotemark{f}\\
2002hw    & Nov. 14.5 & 2592.9  & $V=2594   \pm4$      & $-1\pm4$  & 16.3 & 100 & 750  & 24\\
\enddata 

\tablecomments{Details and references for these data may be found in Appendix A.}
\tablenotetext{a}{Data for $B_{max}$ supplied by Weidong Li, University of California at Berkeley (private communication).}
\tablenotetext{b}{Epoch of observation in days from maximum light in V-band.  ($V_{max}$ is taken to be 2 days after $B_{max}$.)}
\tablenotetext{c}{Estimated magnitude of target in V-band at time of observation.}
\tablenotetext{d}{Total integration time on target.}
\tablenotetext{e}{Signal to noise ratio of combined spectrum before smoothing.}
\tablenotetext{f}{Spectrum obtained in Low Resolution (single prism) mode.}

\end{deluxetable}

\begin{deluxetable}{llrllcccc} 

\tabletypesize{\scriptsize} \tablecolumns{9} \tablewidth{0pc} 
\tablecaption{Estimated Maximum Brightness for Supernovae in the Sample \label{norm_table}} 
\tablehead{\colhead{SN} & \colhead{Date} & \colhead{Epoch\tablenotemark{a}} & \colhead{Host}  & 
\colhead{$v_{rec}$\tablenotemark{b}}& \colhead{E(B-V)\tablenotemark{b} \tablenotemark{c}} & 
\colhead{$V_{max}$ obs.} & \colhead{$M_V$\tablenotemark{d}} & \colhead{$\Delta m_{15}$}}

\startdata
2000dk    & Oct. 2.5  & $+6\pm3$   & NGC 382        & 5228 $\pm27$ & 0.070 & 15.2 $\pm 0.5$ & $-19.5$ & $1.5\pm0.2$\\
2000dm    & Oct. 1.3  & $+1\pm3$   & UGC 11198      & 4507 $\pm44$ & 0.186 & 15.1 $\pm 0.5$ & $-19.7$ & $1.2\pm0.2$\\
2000dn    & Oct. 2.4  & $-6\pm3$   & IC 1468        & 9613 $\pm65$ & 0.046 & 17.0 $\pm0.5$ & $-19.0$ & $0.8\pm0.2$\\
2000do    & Oct. 2.2  & $+4\pm5$   & NGC 6754       & 3257 $\pm10$ & 0.070 & 14.6 $\pm0.5$ & $-19.1$ & $-$\\
2001bf    & May 22.4  & $+5\pm3$   & MGC +04-42-022 & 4647 $\pm31$ & 0.098 & 14.8 $\pm0.5$ & $-19.8$ & $0.8\pm0.2$\\
2001bg    & May 23.3  & $+10\pm2$  & NGC 2608       & 2135 $\pm8$  & 0.039 & 13.7 $\pm0.5$ & $-19.4$ & $1.0\pm0.2$\\
2001br    & May 23.7  & $-2\pm3$   & UGC 11260      & 6184 $\pm48$ & 0.064 & 16.0 $\pm0.5$ & $-19.1$ & $0.9\pm0.2$\\
2001dl    & Aug. 12.4 & $-3\pm3$   & UGC 11725      & 6204 $\pm5$  & 0.053 & 16.0 $\pm0.5$ & $-19.1$ & $0.8\pm0.2$\\
2001en\_1 & Oct. 8.3  & $-4\pm2$   & NGC 523        & 4758 $\pm4$  & 0.054 & 15.0 $\pm0.2$ & $-19.5$ & $1.1\pm0.2$\\
2001en\_2 & Oct. 30.3 & $+18\pm2$   & NGC 523        & 4758 $\pm4$  & 0.054 & 15.0 $\pm0.2$ & -19.5 & $1.1\pm0.2$\\
2002fk    & Sept. 19.5 & $-13\pm2$  & NGC 1309       & 2135 $\pm5$  & 0.040 & 13.2 $\pm0.5$ & -19.5 & $1.1\pm0.2$\\
2002ha    & Nov. 14.3  & $+10\pm4$  & NGC 6962       & 4211 $\pm6$  & 0.098 & 14.7 $\pm0.5$ & -19.6 & $-$\\
2002hw    & Nov. 14.5  & $-1\pm4$   & UGC 52         & 5257 $\pm2$  & 0.111 & 16.2 $\pm0.5$ & -18.7\tablenotemark{e} & $-$\\
\enddata 

\tablenotetext{a}{Epoch of observation in days from maximum light in V-band.}
\tablenotetext{b}{Data from the NASA/IPAC Extragalactic Database.}
\tablenotetext{c}{Galatic extinction.}
\tablenotetext{d}{Using $H_0=65 kms^{-1}Mpc^{-1}$, $A_V = 3.1*E(B-V)$.}
\tablenotetext{e}{Matheson, Challis \& Kirshner report probable extinction due to dust (2002, IAUC 8015)}

\end{deluxetable} 

\begin{deluxetable}{lrrrrrrrr}
\tabletypesize{\scriptsize} \tablecolumns{9} \tablewidth{0pc}
\tablecaption{Expansion velocities in \kms
\label{v_table}} \tablehead{\colhead{SN} &
\colhead{Epoch\tablenotemark{a}} &
\colhead{$\Delta m_{15}$} &
\colhead{\ca} &
\colhead{\mn\tablenotemark{c}} &
\colhead{\mg} &
\colhead{\mg} &
\colhead{\ion{Fe}{2}-edge} &
\colhead{\ion{Si}{2}}\\
\colhead{} & \colhead{(days)} &
 &
\colhead{(1.268 \mum)\tablenotemark{b}} &
\colhead{(0.944 \mum)\tablenotemark{b}} &
\colhead{(0.922 \mum)\tablenotemark{b}} &
\colhead{(1.091 \mum)\tablenotemark{b}} &
\colhead{($\approx 1.57$ \mum)\tablenotemark{b}} &
\colhead{(1.691 \mum)\tablenotemark{b}} }

\startdata
2002fk    &$-13d\pm3$  & $1.1\pm0.2$  & $-$             &  $-$            & $10,400\pm730$ & $11,000\pm620$  & $-$             & $13,500\pm1330$ \\
2000dn    & $-6d\pm2$  & $0.8\pm0.2$  & $-$             &  $-$            & $13,350\pm490$ & $12,650\pm620$  & $-$             & $-$             \\
2000en\_1 & $-4d\pm1$  & $1.1\pm0.2$  & $-$             &  $-$            & $15,050\pm490$ & $15,100\pm620$  & $-$             & $16,150\pm1330$ \\
2001dl    & $-3d\pm2$  & $0.8\pm0.2$  & $-$             &  $-$            & $-$            & $-$             & $-$             & $-$             \\
2001br    & $-2d\pm2$  & $0.9\pm0.2$  & $15,200\pm1770$ &  $-$            & $14,550\pm490$ & $15,300\pm1020$ & $-$             & $12,800\pm1600$ \\
2002hw    & $-1d\pm3$  & $-$          & $13,400\pm890$  &  $-$            & $12,300\pm490$ & $12,450\pm620$  & $-$             & $11,800\pm1330$ \\
2000dm    & $+1d\pm2$  & $1.2\pm0.2$  & $13,400\pm890$  &  $-$            & $11,600\pm730$ & $12,650\pm620$  & $-$             & $13,450\pm1330$ \\
2000do    & $+4d\pm5$  & $-$          & $9,000\pm1770$  &  $11,100\pm950$ & $-$            & $-$             & $10,500\pm1430$ & $-$             \\
2001bf    & $+5d\pm3$  & $0.8\pm0.2$  & $8,100\pm890$   &  $8,050\pm1400$ & $-$            & $-$             & $10,200\pm1430$ & $-$             \\
2000dk    & $+6d\pm3$  & $1.5\pm0.2$  & $7,200\pm1700$  &  $6,850\pm1400$ & $-$            & $-$             & $10,500\pm1430$ & $-$             \\
2001bg    & $+10d\pm2$ & $1.0\pm0.2$  & $-$             &  $-$            & $-$            & $-$             & $10,800\pm1430$ & $-$             \\
2002ha    & $+10d\pm4$ & $-$          & $-$             &  $6,350\pm950$  & $-$            & $-$             & $11,200\pm1430$ & $-$             \\
2001en\_2 & $+18d\pm2$ & $1.1\pm0.2$  & $-$             &  $-$            & $-$            & $-$             & $ 9,800\pm1430$ & $-$             \\

\enddata

\tablenotetext{a}{Epoch of observation in days from maximum light in V-band.}
\tablenotetext{b}{Vacuum wavelength.}
\tablenotetext{c}{The Maganese detection is suggestive but not definitive.}
\end{deluxetable}

\clearpage
\parindent=0pt
\noindent {\bf Figure Captions:}

Fig. 1: {Near infrared spectra of Type Ia supernovae obtained at the
IRTF. The ordinate is log flux and the spectra have been shifted by a
constant.  The abscissa is wavelength in microns.  The spectra are
labeled with the epoch (in days) relative to maximum light (in V) with
the earliest spectrum at the top.  The resolution for each spectrum
after smoothing is given in the top right corner.  Labels for each SN
are have the same order and color as the spectra. }

Fig. 2: {Velocity (solid) and density (dotted) as a function of mass
(in $M_{Ch}$) for the delayed detonation model used as reference
(5p028z22.25; from H\"oflich et al. 2002).}

Fig.3: {Abundances of stable isotopes as a function of expansion
velocity for our reference model (see Fig. \ref{mass_vel}). The ordinate is
log of the mass fraction.  The curves with the highest abundance found
close to the center of the supernova ($< 3000$ \kms) correspond to
$^{54}$Fe, $^{58}$Ni, and $^{56}$Fe.}

Fig. 4: {Evolution of NIR synthetic spectra from the reference model
at three epochs: maximum light in B, at one week, and at two weeks
later.  The fluxes are  normalized at $1.2 \mu m$, and the spectra
are shifted by 2 and 4 units at +1 and +2 weeks, respectively.  As
reference, the V light curve and the evolution of B-V are given in
the right panels.}

Fig. 5: {Log plot of seven spectra from pre-maximum to +1 day (see Section
~\ref{early}).  The ordinate is log flux and the spectra have been
shifted by a constant.  The abscissa is wavelength in microns.  Each
spectrum is labeled with the epoch (in days) relative to maximum light
in V.  Only wavelengths from 0.80-1.80 \mum\ are displayed to
facilitate the identification of features in this region.   The
resolution for each spectrum after smoothing is noted in the top of
the figure.  Labels for each SN are in the same order and color as the
spectra.  Complete spectra (0.80-2.5 \mum) are found in
Figure~\ref{sn13}. }

Fig. 6:  {Log plot of six spectra from +4 to +18 days after $V_{max}$
(see Section~\ref{late}).  The ordinate is log flux and the spectra
have been shifted by a constant.  The abscissa is wavelength in
microns.  The spectra are labeled with the epoch (in days) relative to
maximum light in V.   The resolution for each spectrum after smoothing
is noted in the top of the figure.  Labels for each SN are in the same
order and color as the spectra.  The full sample of ten spectra are
found in Figure~\ref{sn13}. }

Fig. 7: {Doppler velocities of all SNe Ia combined are plotted against
epoch for lines of \mn, \ca, \mn, and the edge of the iron group
feature discussed in Section~\ref{late}.  The abscissa is epoch in
days from maximum light in V. The ordinate is expansion velocity in
$10^3$ \kms\ (See Table~\ref{v_table}).
It can be seen that the measured velocities for \mg\ and \ca\ remain
above 12,000 \kms\ through +1 day after maximum light even though
photospheric velocity has dropped below 10,000 \kms at this time.  The
Edge of the region of incomplete silicon burning (labeled ``Fe edge''
in the figure) also remains above 10,000 \kms\ through +18 days while
the photosphere is less than 5,000 \kms at this time.  Contrast the
\mn\ velocity which seems to follow the photosphere.

\clearpage

\begin{figure}
\plotone{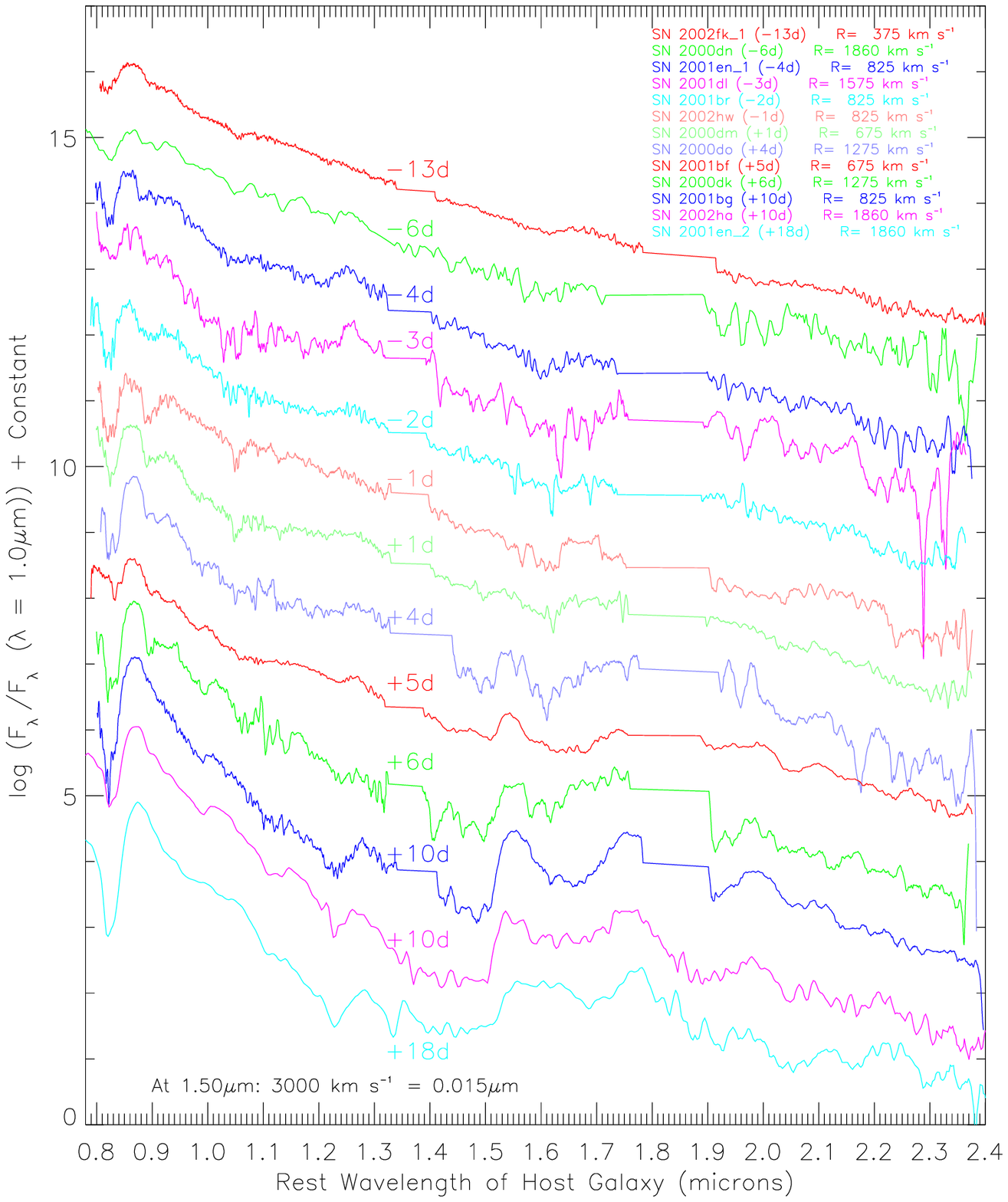}
\caption{ \label{sn13}}
\end{figure}

\begin{figure}
\plotone{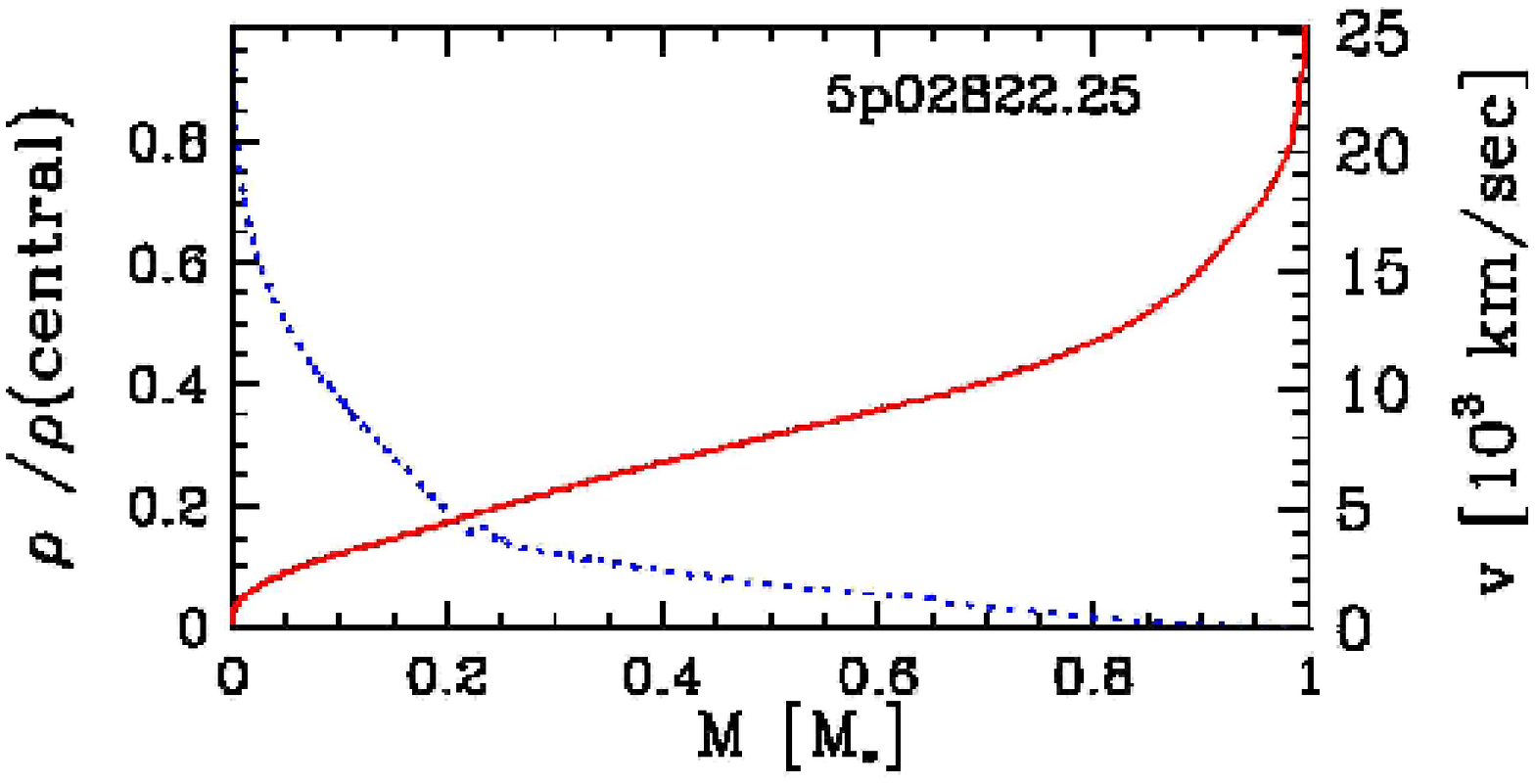}
\caption{ \label{mass_vel}}
\end{figure}

\begin{figure}
\plotone{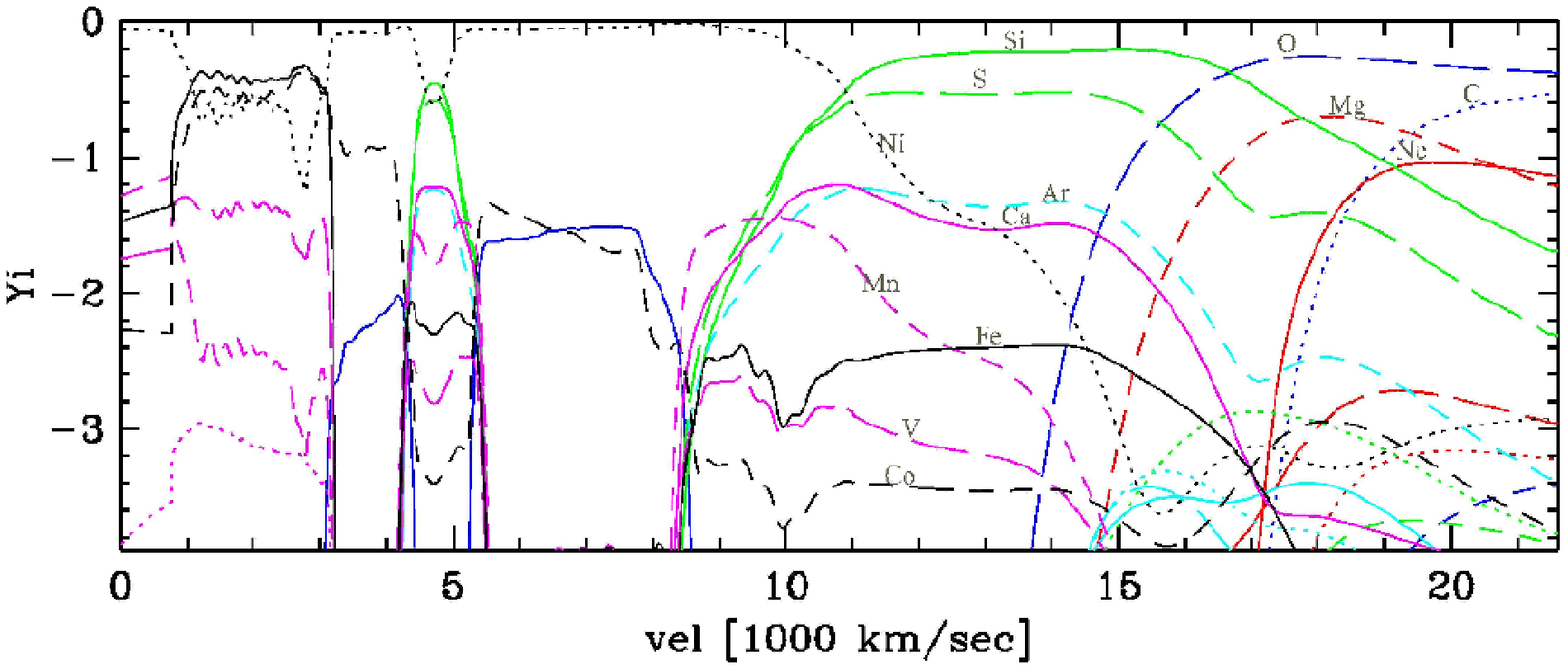}
\caption{ \label{logyi}}
\end{figure}

\begin{figure}
\plotone{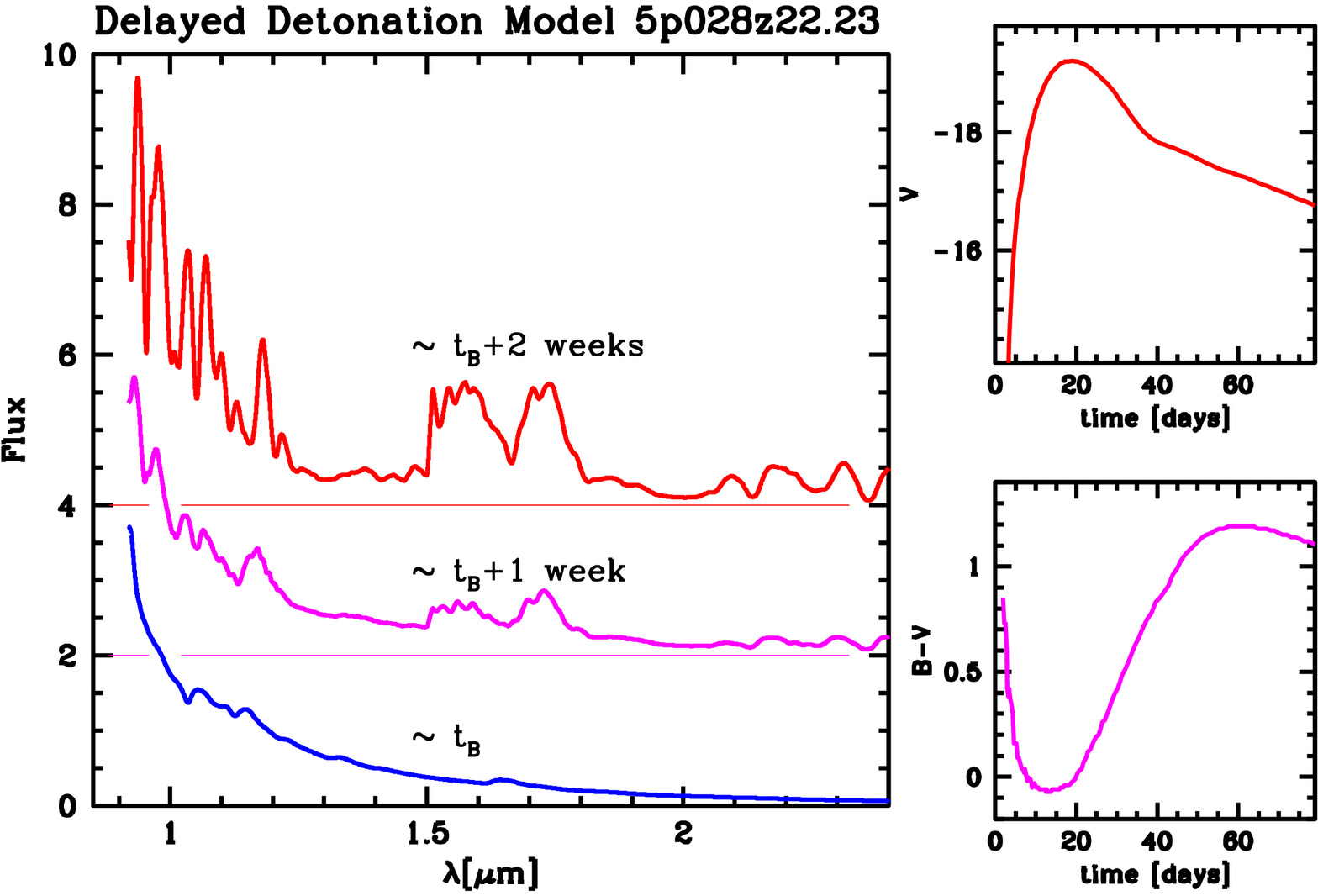}
\caption{ \label{syn_spec}}
\end{figure}

\begin{figure}
\plotone{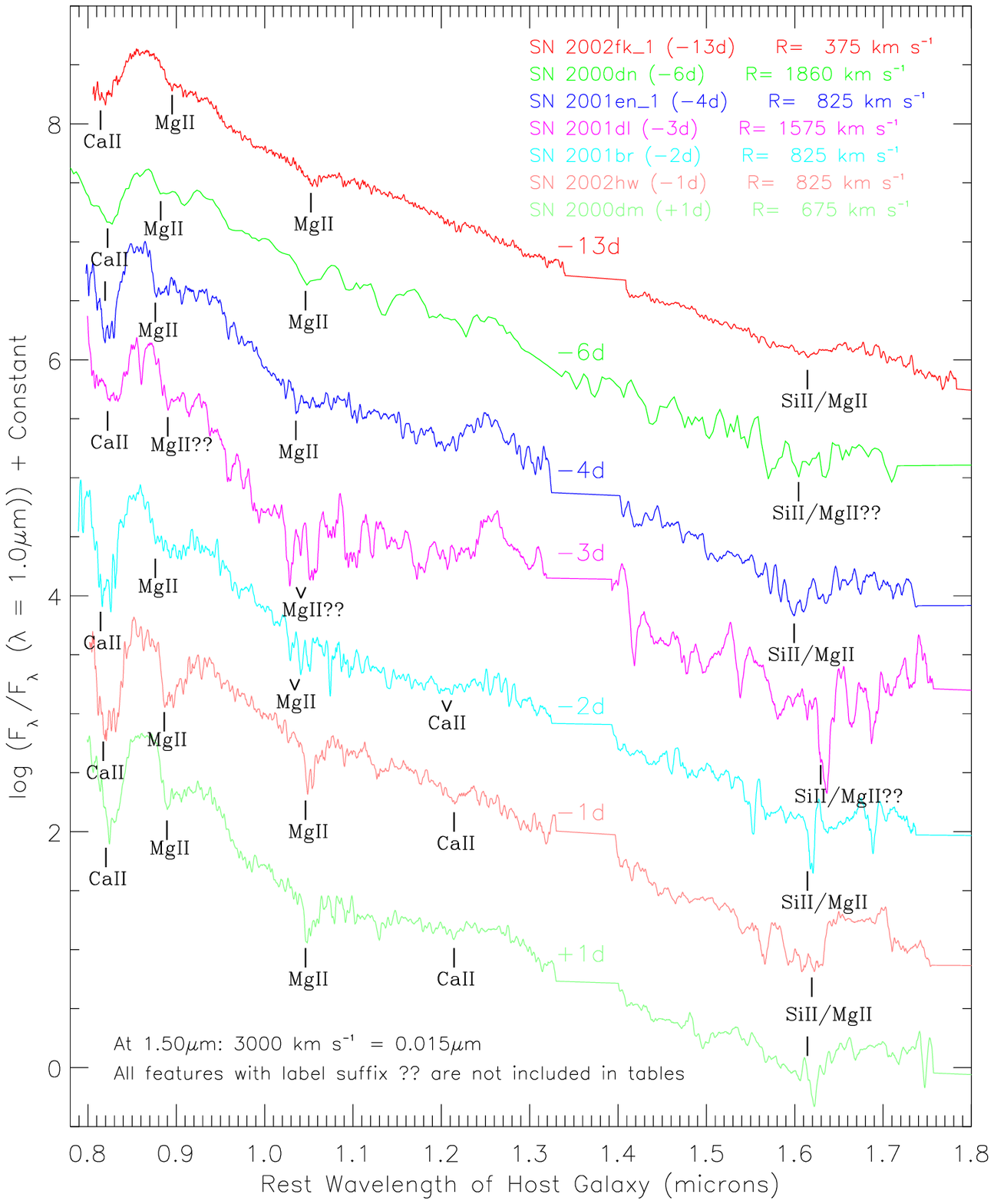}
\caption{ \label{e7}}
\end{figure}

\begin{figure}
\plotone{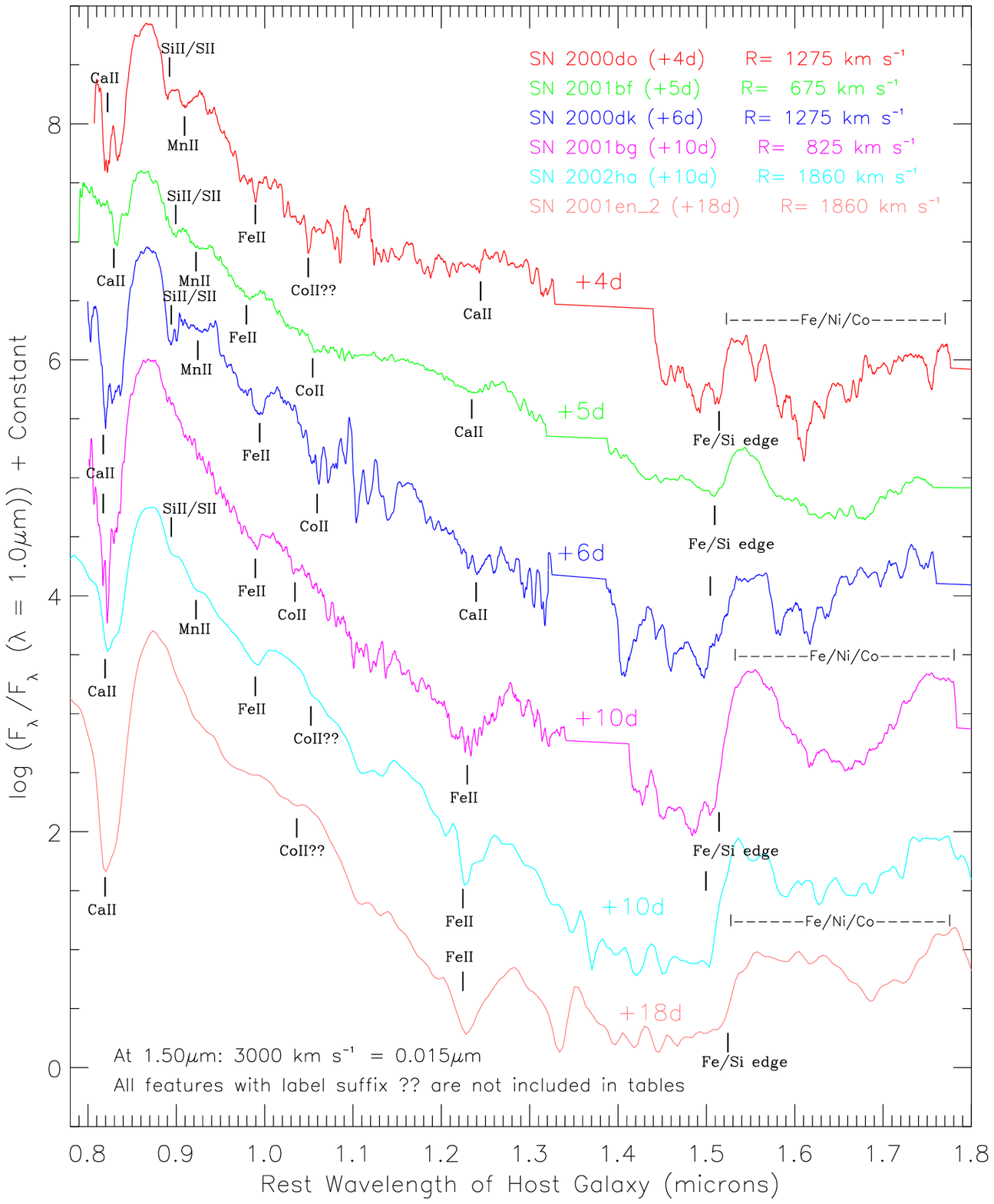}
\caption{ \label{l6}}
\end{figure}

\begin{figure}
\plotone{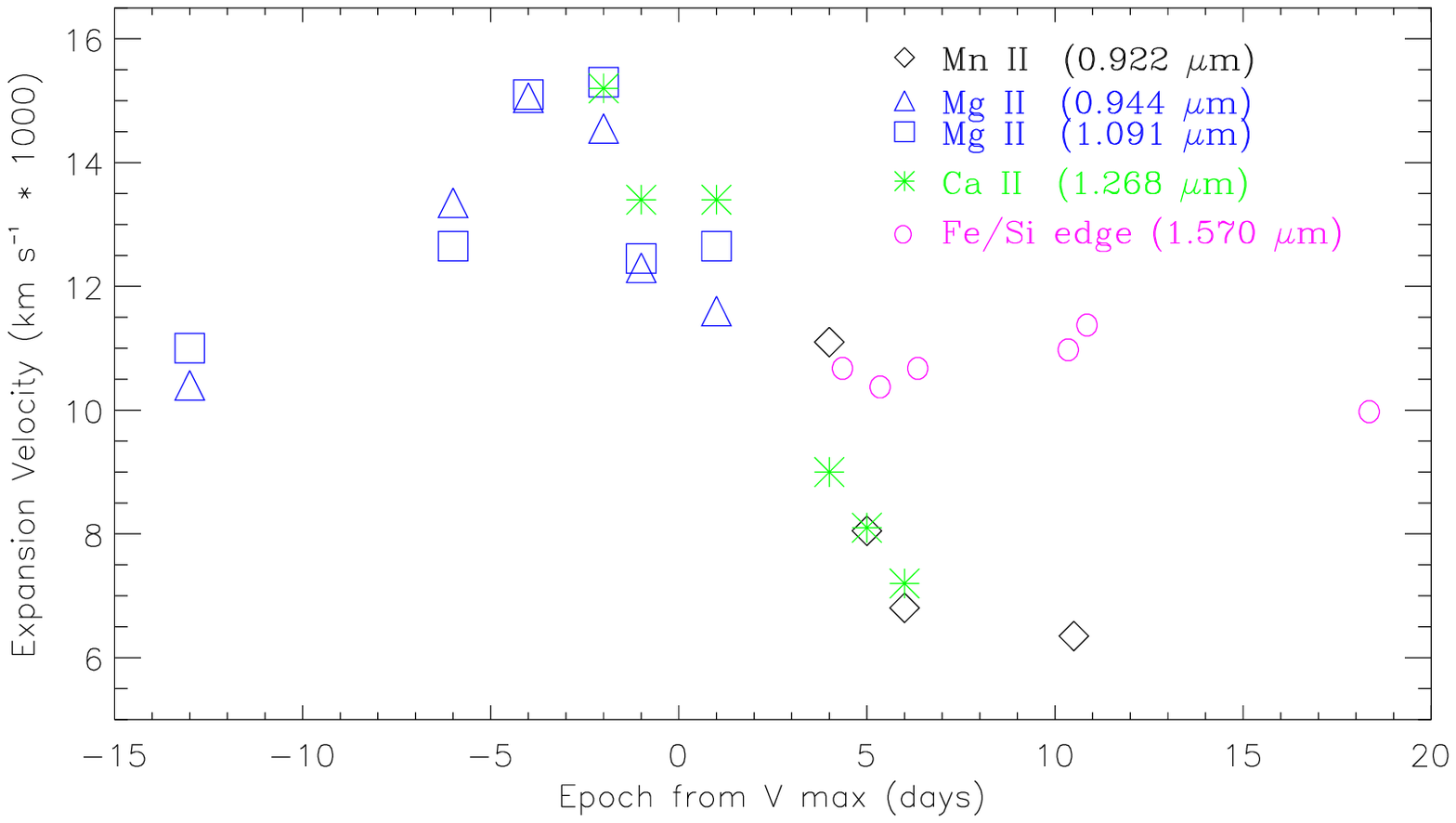}
\caption{ \label{v_plot}}
\end{figure}

\end{document}